\documentclass[fleqn,usenatbib]{mnras}

\usepackage[T1]{fontenc}

\DeclareRobustCommand{\VAN}[3]{#2}
\let\VANthebibliography\thebibliography
\def\thebibliography{\DeclareRobustCommand{\VAN}[3]{##3}\VANthebibliography}

\usepackage{graphicx}
\usepackage{xspace}
\usepackage{subcaption}
\usepackage{etoolbox} % To send emails.
\usepackage{xcolor}   % To define colors.
\usepackage{textgreek}
\usepackage{upgreek}
\let\oldtextsigma\textsigma
\renewcommand{\textsigma}{\oldtextsigma\xspace}
\newcommand{\orcid}[2]{\href{http://orcid.org/#2}{#1}}
\def\h4{\ensuremath{h_4}\xspace}
\newcommand{\mstar}{\ensuremath{M_\star}\xspace}
\newcommand{\msun}{\ensuremath{\mathrm{M_\odot}}\xspace}
\newcommand{\re}{\ensuremath{R_\mathrm{e}}\xspace}
\newcommand{\vse}{\ensuremath{(V/\sigma)_\mathrm{e}}\xspace}
\newcommand{\sigap}{\ensuremath{\sigma_\mathrm{ap}}\xspace}
\newcommand{\kms}{\ensuremath{\mathrm{km\,s^{-1}}}\xspace}

\defcitealias{deugenio+2023}{DE23}

\makeatletter
\newcommand\sendemail[3]{%                %\newcommand\tpj@compose@mailto[3]{%
\edef\@tempa{mailto:#1?subject=#2 }%
\edef\@tempb{\expandafter\html@spaces\@tempa\@empty}%
\href{\@tempb}{#3}}

\catcode\%=11
\def\html@spaces#1 #2{#1%20\ifx#2\@empty\else\expandafter\html@spaces\fi#2}
\catcode\%=14
\makeatother
\newcommand{\ppxf}{{\sc ppxf}\xspace}

\title[Cosmic evolution of $h_4$]{Evolution in the orbital structure of quiescent galaxies from MAGPI, LEGA-C and SAMI surveys: direct evidence for merger-driven growth over the last 7 Gyr}
\author[\orcid{F. D'Eugenio}{0000-0003-2388-8172}~et al.]{\parbox{\textwidth}{
\orcid{Francesco D'Eugenio}{0000-0003-2388-8172}$^{\hyperlink{aff1}{1},\hyperlink{aff2}{2},\hyperlink{aff3}{3}}$\thanks{E-mail: francesco.deugenio@gmail.com},
\orcid{Arjen van der Wel}{0000-0002-5027-0135}$^{\hyperlink{aff3}{3}}$,
\orcid{Joanna M. Piotrowska}{0000-0003-1661-2338}$^{\hyperlink{aff1}{1},\hyperlink{aff2}{2}}$,
\orcid{Rachel Bezanson}{0000-0001-5063-8254}$^{\hyperlink{aff4}{4}}$,
\orcid{Edward N. Taylor}{0000-0002-5522-9107}$^{\hyperlink{aff5}{5}}$,
\orcid{Jesse van de Sande}{0000-0003-2552-0021}$^{\hyperlink{aff6}{6},\hyperlink{aff7}{7}}$,
\orcid{William M. Baker}{0000-0003-0215-1104}$^{\hyperlink{aff1}{1},\hyperlink{aff2}{2}}$,
\orcid{Eric F. Bell}{0000-0002-5564-9873}$^{\hyperlink{aff8}{8}}$,
\orcid{Sabine Bellstedt}{0000-0003-4169-9738}$^{\hyperlink{aff13}{13}}$,
\orcid{Joss Bland-Hawthorn}{0000-0001-7516-4016}$^{\hyperlink{aff6}{6},\hyperlink{aff7}{7}}$,
\orcid{Asa F. L. Bluck}{0000-0001-6395-4504}$^{\hyperlink{aff9}{9}}$,
\orcid{Sarah Brough}{0000-0002-9796-1363}$^{\hyperlink{aff10}{10}}$,
\orcid{Julia J. Bryant}{0000-0003-1627-9301}$^{\hyperlink{aff6}{6},\hyperlink{aff11}{11},\hyperlink{aff7}{7}}$,
\orcid{Matthew Colless}{0000-0001-9552-8075}$^{\hyperlink{aff12}{12},\hyperlink{aff7}{7}}$,
\orcid{Luca Cortese}{0000-0002-7422-9823}$^{\hyperlink{aff13}{13},\hyperlink{aff7}{7}}$,
\orcid{Scott M. Croom}{0000-0003-2880-9197}$^{\hyperlink{aff6}{6},\hyperlink{aff7}{7}}$,
\orcid{Caro Derkenne}{0000-0003-3474-3542}$^{\hyperlink{aff14}{14},\hyperlink{aff7}{7}}$,
\orcid{Pieter van Dokkum}{0000-0002-8282-9888}$^{\hyperlink{aff15}{15}}$,
\orcid{Deanne Fisher}{0000-0003-0645-5260}$^{\hyperlink{aff5}{5},\hyperlink{aff7}{7}}$,
\orcid{Caroline Foster}{0000-0003-0247-1204}$^{\hyperlink{aff10}{10},\hyperlink{aff7}{7}}$,
\orcid{Anna Gallazzi}{0000-0002-9656-1800}$^{\hyperlink{aff16}{16}}$,
\orcid{Anna de~Graaff}{0000-0002-2380-9801}$^{\hyperlink{aff17}{17},\hyperlink{aff18}{18}}$,
\orcid{Brent Groves}{0000-0002-9768-0246}$^{\hyperlink{aff13}{13}}$,
\orcid{Josha van Houdt}{0000-0003-1888-3705}$^{\hyperlink{aff18}{18}}$,
\orcid{Claudia del P. Lagos}{0000-0003-3021-8564}$^{\hyperlink{aff13}{13},\hyperlink{aff7}{7}}$,
\orcid{Tobias J. Looser}{}$^{\hyperlink{aff1}{1},\hyperlink{aff2}{2}}$,
\orcid{Roberto Maiolino}{0000-0002-4985-3819}$^{\hyperlink{aff1}{1},\hyperlink{aff2}{2},\hyperlink{aff19}{19}}$,
\orcid{Michael Maseda}{0000-0003-0695-4414}$^{\hyperlink{aff20}{20}}$,
\orcid{J. Trevor Mendel}{0000-0002-6327-9147}$^{\hyperlink{aff12}{12},\hyperlink{aff7}{7}}$,
\orcid{Angelos Nersesian}{0000-0001-6843-409X}$^{\hyperlink{aff3}{3}}$,
\orcid{Camilla Pacifici}{0000-0003-4196-0617}$^{\hyperlink{aff21}{21}}$,
\orcid{Adriano Poci}{0000-0003-4196-0617}$^{\hyperlink{aff22}{22},\hyperlink{aff14}{14}}$,
\href{https://www.usm.uni-muenchen.de/~rhea/index.html}{Rhea-Silvia Remus}$^{\hyperlink{aff23}{23}}$,
\orcid{Sarah M. Sweet}{0000-0002-1576-2505}$^{\hyperlink{aff24}{24}}$,
\orcid{Sabine Thater}{0000-0003-1820-2041}$^{\hyperlink{aff25}{25}}$,
\orcid{Kim-Vy Tran}{0000-0001-9208-2143}$^{\hyperlink{aff10}{10},\hyperlink{aff7}{7},\hyperlink{aff26}{26}}$,
\orcid{Hannah {\"U}bler}{0000-0003-4891-0794}$^{\hyperlink{aff1}{1},\hyperlink{aff2}{2}}$,
 \orcid{Lucas M. Valenzuela}{0000-0002-7972-9675}$^{\hyperlink{aff23}{23}}$,
\orcid{Emily Wisnioski}{0000-0003-1657-7878}$^{\hyperlink{aff12}{12},\hyperlink{aff7}{7}}$,
\orcid{Stefano Zibetti}{0000-0003-1734-8356}$^{\hyperlink{aff16}{16}}$
}
\vspace{0.4cm}
\\
\parbox{\textwidth}{
\hypertarget{aff1}{$^{1}$}Kavli Institute for Cosmology, University of Cambridge, Madingley Road, Cambridge, CB3 0HA, United Kingdom\\
\hypertarget{aff2}{$^{2}$}Cavendish Laboratory - Astrophysics Group, University of Cambridge, 19 JJ Thomson Avenue, Cambridge, CB3 0HE, United Kingdom\\
\hypertarget{aff3}{$^{3}$}Sterrenkundig Observatorium, Universiteit Gent, Krijgslaan 281 S9, B-9000 Gent, Belgium\\
\hypertarget{aff4}{$^{4}$}Department of Physics and Astronomy and PITT PACC, University of Pittsburgh, Pittsburgh, PA 15260, USA\\
\hypertarget{aff5}{$^{5}$}Centre for Astrophysics and Supercomputing, Swinburne University of Technology, Hawthorn, VIC 3122, Australia\\
\hypertarget{aff6}{$^{6}$}Sydney Institute for Astronomy, School of Physics, The University of Sydney, NSW, 2006, Australia\\
\hypertarget{aff7}{$^{7}$}ARC Centre of Excellence for All Sky Astrophysics in 3 Dimensions (ASTRO 3D), Australia\\
\hypertarget{aff8}{$^{8}$}Department of Astronomy, University of Michigan, Ann Arbor, MI 48109, USA\\
\hypertarget{aff9}{$^{9}$}Department of Physics, Florida International University, 11200 SW 8th Street, Miami, FL, USA\\
\hypertarget{aff10}{$^{10}$}School of Physics, University of New South Wales, NSW 2052, Australia\\
\bigskip
\emph{\normalsize Remaining affiliations are listed at the end of the paper}
}
}

\date{Accepted XXX. Received YYY; in original form ZZZ}

\pubyear{2023}

\begin{document}
\label{firstpage}
\pagerange{\pageref{firstpage}--\pageref{lastpage}}
\maketitle

% Abstract of the paper
\begin{abstract}
We present the first study of spatially integrated higher-order stellar
kinematics over cosmic time. We use deep rest-frame optical spectroscopy
of quiescent galaxies at redshifts $z=0.05$, 0.3 and 0.8 from the SAMI,
MAGPI and LEGA-C surveys to measure the excess kurtosis $h_4$ of the
stellar velocity distribution, the latter parametrised as a Gauss-Hermite series.
Conservatively using a redshift-independent cut in stellar mass
($M_\star = 10^{11}\,\mathrm{M_\odot}$), and matching the stellar-mass
distributions of our samples, we find 7~\textsigma evidence of $h_4$ increasing
with cosmic time, from a median value of $0.019\pm0.002$ at $z=0.8$ to
$0.059\pm0.004$ at $z=0.06$.
Alternatively, we use a physically motivated sample selection, based on the mass
distribution of the progenitors of local quiescent galaxies as inferred from
numerical simulations; in this case, we find 10~\textsigma evidence.
This evolution suggests that, over the last 7~Gyr, there has been a gradual
decrease in the rotation-to-dispersion ratio and an increase in the radial
anisotropy of the stellar velocity distribution, qualitatively consistent 
with accretion of gas-poor satellites. These findings demonstrate that massive
galaxies continue to accrete mass and increase their dispersion support after
becoming quiescent.
\end{abstract}

% Select between one and six entries from the list of approved keywords.
% Don't make up new ones.

%%%%%%%%%%%%%%%%%%%%%%%%%%%%%%%%%%%%%%%%%%%%%%%%%%

%%%%%%%%%%%%%%%%% BODY OF PAPER %%%%%%%%%%%%%%%%%%

\section{Introduction}

The most massive galaxies in the Universe are thought to form in two phases. The
first stage is dominated by dissipative gas accretion and \textit{in-situ} star
formation. In the second stage, after cosmic noon \citep{madau+dickinson2014,
forsterschreiber+wuyts2020}, massive galaxies typically become quiescent, but
continue to grow in both mass and size through accretion of low-mass
gas-poor satellites \citep{bezanson+2009, naab+2009, oser+2010, oser+2012,
stockmann+2021}. %, naab+2014}.

This theoretical picture was drawn to explain the observed changes in the
average properties of quiescent galaxies across cosmic time.
Observing campaigns of the nearby Universe have revealed that local massive
quiescent galaxies are dispersion dominated \citep{davies+1983, emsellem+2011}
and intrinsically round or triaxial \citep{lambas+1992, vincent+ryden2005,
weijmans+2014, foster+2017, hongyu+2018}; they have a large fraction of dynamically warm
and hot orbits \citep{zhu+2018, santucci+2022}, and flat or
even `u'-shaped stellar-age radial profiles \citep{zibetti+2020}.
In contrast, massive quiescent galaxies in the early Universe were smaller
\citetext{\citealp{daddi+2005a}, \citealp{trujillo+2007}, which suggests size
evolution}, and intrinsically flatter \citetext{\citealp{vanderwel+2011},
\citealp{chang+2013}, suggesting a higher rotation-to-dispersion
ratio compared to local quiescent galaxies, as confirmed by
e.g. \citealp{newman+2015}}.

However, connecting primordial to local galaxies is complicated by progenitor
bias \citep{vandokkum+franx2001}: the progenitors of some of the local
quiescent galaxies were not already quiescent several billion years ago.
This means that, in principle, the physical differences between local and
distant quiescent galaxies could be due entirely to changing demographics.

Indeed, there is now overwhelming evidence for inside-out growth of star-forming
galaxies, both in the local Universe \citep[where we can even measure the
instantaneous size-growth rate,][]{pezzulli+2015} and at all epochs until
cosmic noon \citep{robotham+2022, wang+2019, nelson+2016, paulino-afonso+2017, suzuki+2019}.
At any given moment in the history of the Universe, star-forming galaxies are on
average larger than quiescent galaxies of the same stellar mass
\citep[e.g.][]{vanderwel+2014, mowla+2019}; for this reason, also newly quiescent
galaxies have larger average size compared to the extant quiescent
population of the same mass \citep{newman+2012, carollo+2013, wu+2018}, potentially
explaining the increase in the average physical size of the quiescent galaxy
population over cosmic time.\footnote{In principle, star-forming
galaxies could also experience sudden structural changes just before becoming
quiescent, for example as a result of a final, central starburst
\citep[e.g.]{chen+2019, deugenio+2020}, but this `rapid path to quiescence'
\citep{wu+2018} does not necessarily lead to changes in size \citetext{cf.
\citealp{deugenio+2020}, \citealp{wu+2020}, and \citealp{setton+2020, setton+2022}}.
Besides, it is a rare evolutionary path in the local Universe \citep{rowlands+2018} and,
even around cosmic noon, when it is most common, it seem to explain only half
the observed growth in the comoving number density of the quiescent population
\citep{belli+2015}.}
To account for the effect of this progenitor bias on the size evolution
of quiescent galaxies, one must keep track of the properties of both quiescent
\textit{and} star-forming galaxies.

However, large photometric and grism-spectroscopy surveys such as CANDELS
\citep{grogin+2011, koekemoer+2011} and 3D-HST \citep{brammer+2012} have
unequivocally shown that demographic changes cannot explain, alone, the observed
size difference between early and contemporary quiescent galaxies. The comoving
volume density of compact quiescent galaxies \emph{decreases} with cosmic time,
which requires physical growth of individual galaxies \textit{after} they became
quiescent \citetext{\citealp{vanderwel+2014}; confirming earlier indications
from \citealp{taylor+2010}}.

It is to explain this inferred size growth that minor dry mergers
were first invoked \citep{bezanson+2009, naab+2009}. Alternative explanations do
not account for all the observations. Star-formation episodes after quiescence
(rejuvenation) can be ruled out (as main mechanism) based on direct measurements
of the star-formation
history of quiescent galaxies \citetext{\citealp{chauke+2019}; besides, this
mechanism is not consistent with the observed changes in shape, e.g.
\citealp{vanderwel+2011}}. While major dry mergers could in principle explain the
observed evolution, their predicted rate \citep{oser+2012, nipoti+2012,
sweet+2017} appears insufficient to account for the magnitude of the observed
changes, because of the linear relation between mass and size growth in major
mergers \citep{naab+2009}.
In addition, the small number of major dry mergers may not reproduce the
intrinsic shape and the spatially resolved stellar kinematics of the most
massive galaxies, because the orbital angular momentum of the progenitors is
locked within the stellar orbits of the remnant \citetext{\citealp{bois+2011},
but see e.g. \citealp{taranu+2013} and \citealp{lagos+2022} for a contrasting
view}.

In contrast, minor dry mergers are consistent with all the observed
changes. They explain the observed change in shape and the loss of angular
momentum, while the low relative mass of the satellites
\citep[mass ratio 6:1 and higher,][]{naab+2014}, explains the
steep radial-to-mass growth rate \citep{bezanson+2009, naab+2009}.
The conservation of orbital
energy between the accreted satellite and its stars means that most stars are
dispersed along radially anisotropic orbits, changing the light distribution and stellar
population content more at large radii than at small radii, as required by
observations of weak evolution in the central surface mass density
\citep[e.g.][]{bezanson+2009}.

Unfortunately, by definition, minor mergers are hard to constrain
observationally beyond the local Universe. In particular, the uncertainty on the
timescale over which a merger is recognisable translates into large
uncertainties on the merger rate
\citep{newman+2012}. However, a sufficient number of minor mergers will have a visible
impact on the stellar kinematics of the accreting galaxy.
In particular,
if accreted stars are dispersed about the orbit of the satellite, we
expect them to move along elongated orbits, which is different from the results of
both star-formation and major dry mergers \citep{bois+2011}. By comparing the
velocity distribution of massive quiescent galaxies across cosmic time, we can
test another prediction of the minor-dry-merger hypothesis.

While integral field spectroscopy surveys enable us to accurately model the
stellar orbital distribution of local galaxies \citep{cappellari+2007, zhu+2018}
and to compare the detailed properties of their spatially resolved velocity
distributions to simulations \citep{vandesande+2017a, vandesande+2019}, this
type of observation remains out of reach beyond the local Universe, where large
samples with high-quality measurements are limited to integrated spectra.
Fortunately, even spatially integrated measurements preserve
some information about the assembly history of galaxies. There are some caveats
to this statement: following a major merger, the distribution function does relax
thus leading to loss of information \citep[e.g.][]{lynden-bell1967}. However, in
general, relative to an isotropic stellar system, an over-abundance of radial
orbits is reflected in the shape of the stellar velocity distribution, causing it
to deviate from a Gaussian and become more peaked with more prominent wings
(leptokurtic); conversely, an over-abundance of circular orbits reflects
a less-peaked (platykurtic) velocity distribution \citep{vandermarel+franx1993}.
So to test the hypothesis that the observed evolution in the kinematics and
size of massive, quiescent galaxies is due to the cumulative effect of many
minor dry mergers, we need high-quality integrated spectra for a statistical
sample of galaxies spanning a sizeable fraction of the history of the Universe.

Until recently, the necessary combination of large sample size and high-quality
integrated spectra did not exist, but the advent of large, absorption-line
spectroscopy surveys of the early Universe changed this state of affairs.

In this work, we leverage the extraordinary quality, sample size and large
look-back time of the LEGA-C and MAGPI data, complemented with local
observations from the SAMI Galaxy Survey, to investigate the cosmic evolution
of the excess kurtosis \citep[parametrised by \h4,][]{vandermarel+franx1993,
gerhard1993} as a direct tracer of the assembly history of galaxies.
After introducing the data and sample in \S~\ref{s.das}, we show that \h4
increases with cosmic time (\S~\ref{s.r.ss.h4z}) and discuss the implications
of our
findings on the size growth of quiescent galaxies (\S~\ref{s.d}). A summary of
our findings is provided in \S~\ref{s.c}.

Throughout this article, we use a flat $\Lambda$CDM cosmology with $H_0 = 70 \;
\kms \; \mathrm{Mpc^{-1}}$ and $\Omega_m = 0.3$. 
All stellar mass measurements assume a Chabrier initial mass function
\citep{chabrier2003}.

\section{Data}\label{s.das}

In this section, we start by presenting the data (\S~\ref{s.das.ss.ds}), which
we draw from three different surveys: the SAMI Galaxy Survey (redshift $z \approx 0$,
\S~\ref{s.das.ss.ds.sss.sami}), the MAGPI survey
($z \approx 0.3$, \S~\ref{s.das.ss.ds.sss.magpi}), and the LEGA-C survey
($z \approx 0.7$, \S~\ref{s.das.ss.ds.sss.legac}). We then explain
how this heterogeneous dataset is homogenised (\S~\ref{s.das.ss.match}) and how
the resulting one-dimensional (1-d) spectra are used to measure \h4
(\S~\ref{s.das.ss.hok}). Finally, in \S~\ref{s.das.ss.anc}, we describe ancillary
measurements obtained from the literature.

\subsection{Data sources}\label{s.das.ss.ds}

\subsubsection{The SAMI Galaxy Survey}\label{s.das.ss.ds.sss.sami}

The SAMI Galaxy Survey (hereafter simply: SAMI) is a large, integral-field
optical-spectroscopy survey of local galaxies. It spans a range of redshifts
$0.04 < z < 0.095$, a stellar mass range $10^7 < \mstar < 10^{12} \,
\msun$, all morphological types and environments \citetext{from
isolated galaxies to eight clusters, local environment density $0.1 < \Sigma_5 <
100 \, \mathrm{Mpc^{-2}}$, \citealp{bryant+2015}, \citealp{owers+2017}; see
\citealp{brough+2017} for the definition of $\Sigma_5$}.
SAMI observations were obtained at the 3.9-m Anglo-Australian Telescope, using
the Sydney-AAO Multi-object Integral field spectroscopy instrument
\citep[hereafter, the SAMI instrument;][]{croom+2012}.
The SAMI instrument has 13 integral field units (IFUs), consisting of a
fused-fibre bundle \citep[hexabundle;][]{bland-hawthorn+2011, bryant+2014} of
61 individual fibres of 1.6-arcsec diameter, giving a total IFU diameter of 15
arcsec. The 13 IFUs are deployable inside a 1-degree field of view, and are
complemented by 26 individual sky fibres.
The fibres are fed to the double-beam AAOmega spectrograph \citep{sharp+2006};
the blue arm was configured with the 570V grating at 3750--5750~\AA\ ($R=1812$,
$\sigma = 70.3$~\kms) and the red arm was configured with
the R1000 grating at 6300--7400~\AA\ \citep[$R=4263$, $\sigma = 29.9$~\kms
][]{vandesande+2017a}.
Each galaxy was exposed for approximately
3.5~hours, stacking seven 0.5-h dithered exposures \citep{sharp+2015}.
The median seeing full width at half maximum (FWHM) of SAMI is $2.06 \pm
0.40$~arcsec. The data reduction
process is outlined in \citet{sharp+2015} and \citet{allen+2015}. Ensuing 
improvements are described in the public data release papers
\citep{green+2018, scott+2018}. Here we use 3068 unique datacubes from the
third and final public data release \citep[Data Release 3, hereafter: DR3][]{
croom+2021a}.

\subsubsection{MAGPI}\label{s.das.ss.ds.sss.magpi}

The Middle Ages Galaxy Properties with Integral field spectroscopy survey
\citep[hereafter, MAGPI][]{foster+2021} is a Large Program with the Multi-Unit
Spectroscopic Explorer \citep[MUSE,][]{bacon+2010} on the 8-m European Southern 
Observatory (ESO) Very Large Telescope (VLT). MAGPI
aims to study spatially resolved galaxy properties in the uncharted cosmic
`Middle Ages' at $z \approx 0.3$, between the epoch of `classic' local surveys
(like SAMI) and higher-redshift studies (like LEGA-C).
MUSE was configured in the large-field mode ($1\times1$-arcmin$^2$ field of
view), aided by Ground Layer Adaptive Optics GALACSI \citep{arsenault+2008,
strobele+2012} to achieve a spatial resolution with median FWHM of
0.6--0.8~arcsec (comparable, in physical units, to the spatial resolution of
SAMI). MAGPI spectra cover the approximate rest-frame wavelength range
$3600 < \lambda < 7200$~\AA, with a median spectral resolution FWHM of
1.25~\AA\ (inside one effective radius, the FWHM varies by 3~per\ cent).

The sample consists of 60
central galaxies, drawn from the Galaxy and Mass Assembly survey \citep[GAMA;][]{
driver+2011, liske+2015, baldry+2018} and from two legacy programs, targeting
clusters Abell~370 (Program ID 096.A-0710; PI: Bauer) and
Abell~2744 (Program IDs: 095.A-0181 and 096.A-0496; PI: Richard). In addition
to the central galaxies, MAGPI will concurrently observe one hundred satellite
galaxies in the target redshift range, plus any background galaxy inside the
field of view.

The survey is in progress, but MAGPI has already obtained data for fifteen
fields, which we use in this work. An overview of the observations and
data reduction is provided in the survey paper \citep{foster+2021}, while a
complete description of the data reduction pipeline
\citetext{based on the MUSE pipeline, \citealp{weilbacher+2020} and on the
Zurich Atmosphere Purge sky-subtraction software, \citealp{soto+2016}}, will
be provided in an upcoming paper (Mendel~et~al., in~prep.). Each MAGPI cube is
segmented into `minicubes', centred on individual galaxy detections.

\subsubsection{LEGA-C}\label{s.das.ss.ds.sss.legac}

Our redshift baseline is completed by the Large Early Galaxy Astrophysics
Census, a large, deep optical-spectroscopy survey of galaxies between
$0.6 < z < 1.0$ \citep{vanderwel+2016}. The LEGA-C sample consists of
3000 primary galaxies, $K_s$-band selected from the UltraVISTA catalogue
\citep{muzzin+2013a}.

Observations were carried at the ESO VLT using the VIMOS spectrograph \citep{lefevre+2003} in its
multi-object configuration, with mask-cut slits of length $\geq 8$~arcsec and
width 1~arcsec; all slits from the main survey were oriented in the
North-South direction, so they were aligned randomly relative to the major axes
of the target galaxies. The seeing median FWHM, measured from a Moffat fit to
the slit data, is 0.75~arcsec \citep{vanhoudt+2021}. The typical observed-frame
spectral interval spans $6300 < \lambda < 8800$~\AA (the exact range depends on
the slit position inside the mask). The spectral resolution is $R=2500$
\citep[but the effective spectral resolution is $R=3500$, due to the LEGA-C
targets underfilling the slit;][]{straatman+2018}. Each mask was exposed for
20~h, reaching a continuum signal-to-noise ratio $S/N\approx$20~\AA$^{-1}$.
Thanks to the depth of these observations, most targets have successful
kinematics measurements (93~per\ cent), resulting in a mass-completeness limit of
$10^{10.5} \, \msun$ \citep{vanderwel+2021}.

In this work, we use the 1-d LEGA-C spectra from the third public data release
\citep[DR3,][]{vanderwel+2021}. These were obtained from optimal extraction
\citep{horne1986} of the 2-d spectra. The large physical width of the LEGA-C
slits (7.5~kpc at $z=0.8$) means that the 1-d spectra sample
a representative fraction of the targets' light (the ratio between the slit width
and the circularised galaxy diameter is $1.0\pm0.5$ for our redshift-evolution
sample, see \S~\ref{s.samp} for the sample selection). To measure \h4, we use
the method outlined in \S~\ref{s.das.ss.hok} and described in
\citet[][hereafter: \citetalias{deugenio+2023}]{deugenio+2023}. We
set the observed-frame spectral FWHM to a wavelength-independent value of
2.12~\AA\  \citep[corresponding to 86~\kms,][]{vanderwel+2021}.
Even though we use emission-line subtracted spectra \citep{bezanson+2018a}, the
precision and accuracy of the subtraction do not affect our measured kinematics
\citepalias{deugenio+2023}.

LEGA-C is the highest-redshift survey we use, so it has the least spatial
information and narrowest wavelength range. For this reason, in order to
draw a fair comparison with the other two datasets, we match the quality of the
SAMI and MAGPI surveys to reproduce the observing setup of LEGA-C
(\S~\ref{s.das.ss.match}). The impact of the
different observing setup is discussed in \S~\ref{s.r.ss.mbias}.

\subsection{Data homogenisation}\label{s.das.ss.match}

\begin{figure*}
  \includegraphics[type=pdf,ext=.pdf,read=.pdf,width=1.\textwidth]{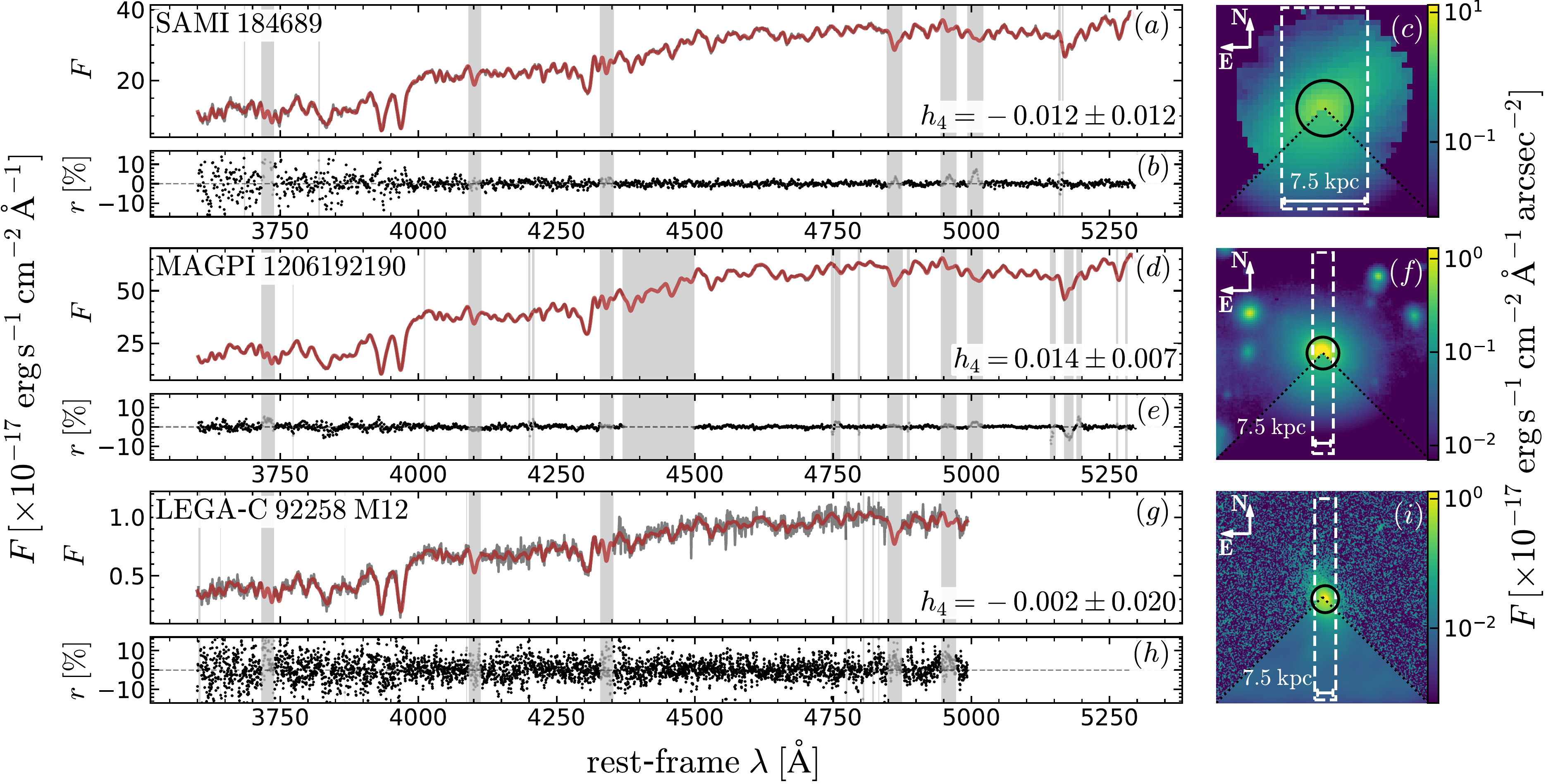}
  {\phantomsubcaption\label{f.das.ppxf.a}
   \phantomsubcaption\label{f.das.ppxf.b}
   \phantomsubcaption\label{f.das.ppxf.c}
   \phantomsubcaption\label{f.das.ppxf.d}
   \phantomsubcaption\label{f.das.ppxf.e}
   \phantomsubcaption\label{f.das.ppxf.f}
   \phantomsubcaption\label{f.das.ppxf.g}
   \phantomsubcaption\label{f.das.ppxf.h}
   \phantomsubcaption\label{f.das.ppxf.i}}
  \caption{Comparison between three galaxies, randomly chosen from the
  SAMI, MAGPI and LEGA-C samples. For each galaxy, we show the data (dark grey)
  and best-fit spectra (red), alongside the relative residuals (black dots).
  Vertical lines/regions are masked because of possible emission lines
  (regardless of whether lines were actually detected) or because of instrument
  setup (e.g. the GALACSI laser band for MAGPI, panels~\subref{f.das.ppxf.d}
  and~\subref{f.das.ppxf.e}). The inset figures show the
  galaxy images (from the datacube for SAMI and MAGPI,
  panels~\subref{f.das.ppxf.c} and~\subref{f.das.ppxf.f}, from HST F814W for
  LEGA-C, panel~\subref{f.das.ppxf.i}).
  In each of the three images (panels~\subref{f.das.ppxf.c}, \subref{f.das.ppxf.f}
  and~\subref{f.das.ppxf.i}), we indicate the galaxy effective radius with a solid
  black circle, and the applied slit with a dashed white line. The lowest quadrant
  of each image shows the data convolved to the ground-based spatial resolution
  of LEGA-C.
  }\label{f.das.ppxf}
\end{figure*}

To match the high-redshift slit spectroscopy of LEGA-C, we artificially `redshift'
the SAMI and MAGPI galaxies to z=0.78. This is done in two steps: \textit{i}) blur
the point-spread function to match the LEGA-C seeing, and \textit{ii}) extract the
spectrum within a LEGA-C-like slit. In the first step, we convolve the datacubes with a
Gaussian kernel; the Gaussian FWHM is calculated for each galaxy as the difference
in quadrature between the median LEGA-C seeing FWHM of 5.6~kpc (at z=0.78) and the
observed SAMI or MAGPI FWHM (for SAMI, this value is obtained from the SAMI DR3
catalogue; for MAGPI, the values are retrieved from the processed datacubes). Note
that we also match the three surveys in wavelength, by removing any data outside
the rest-frame interval 3600--5300~\AA. The matching procedure is illustrated in the right column
of Fig.~\ref{f.das.ppxf}, where the top, middle and bottom rows show data from
SAMI, MAGPI and LEGA-C, respectively.
For MAGPI and SAMI, the images (panels~\subref{f.das.ppxf.c}
and~\subref{f.das.ppxf.f}) are reconstructed from the datacubes, shown at the
original spatial resolution. The bottom quadrant of each panel shows the result
of the convolution to match the LEGA-C seeing. For LEGA-C, the image is the HST
F814W photometry (panel~\subref{f.das.ppxf.i}); the bottom quadrant has been
convolved with the ground-based LEGA-C seeing, to illustrate the VLT/VIMOS view of
the target.
For each galaxy, we calculate the noise spectrum by
applying the square of the kernel to the variance datacube, but we ignore spatial
correlations (we estimate the effect spatial correlation on the noise by rescaling
the noise spectrum after the first fit; see \S~\ref{s.das.ss.hok}). After this
procedure, the datacube matches the average spatial resolution of LEGA-C.

The second step
consists of creating the 2-d slit spectrum; we convolve the datacube with a slit
of width 7.5~kpc (corresponding to 1~arcsec at z=0.78) and length 75~kpc (white
dashed rectangle in right column of Fig.~\ref{f.das.ppxf}; in practice, the slit
always exceeds the size of the SAMI IFU and of most MAGPI minicubes).
The slits are placed on the centre of the galaxy\footnote{To test the effect
of the precision in the slit centering, we created random realisations of the
SAMI spectra with a dispersion-direction slit offset drawn from a normal
distribution with standard deviation 1~pixel \citep{vanhoudt+2021}. These
offsets do not change our results} and are oriented in the North-South direction
(i.e., randomly compared to the position angle of the target).

Next we simulate the
degeneracy between Doppler shift and spatial offset along the dispersion direction.
For each IFU spatial pixel (spaxel), we calculate its spatial offset from the centre
of the mock slit, in units of (equivalent) LEGA-C pixels; because each LEGA-C spaxel
consists of five detector pixels, we calculate this offset for up to $\pm$2.5 pixels.
For LEGA-C, each detector pixel corresponds to 0.6~\AA, so the wavelength shift is in
the range $\pm$1.5~\AA. We apply this shift by re-binning the spectrum of each IFU
spaxel to the new wavelength grid.

For each spectral pixel, the flux in each slit spaxel is calculated as a linear sum
of the flux of all SAMI spaxels that intersect the slit spaxel, weighted only by the
overlapping area fraction. After calculating the spectrum for each slit spaxel, we
optimally extract the resulting 2-d spectrum and obtain the final 1-d spectrum
\citep{horne1986}. This latter step is in principle different from the procedure
adopted in \citet{vanderwel+2021}, who use HST photometry with high spatial
resolution to guide the extraction. But in practice the $S/N$ of SAMI and MAGPI is so
high that using a prior from photometry is not required, and we can use the observed
slit profile for the extraction.
Each spectrum is cut between 3600--5300~\AA, to match the typical rest-frame
wavelength range of LEGA-C.
For the MUSE 1-d spectra, we also mask the wavelength region affected by
the GALACSI laser (large grey shaded area in the middle row of Fig.~\ref{f.das.ppxf}).

One potential concern is to quantify what random/systematic errors in the \h4
measurements are introduced in the process of degrading the SAMI data to match
LEGA-C. To address this concern we use 1-d spectra from an elliptical aperture
with semi-major axis \re (see \S~\ref{s.das.ss.anc}).
These spectra provide a `baseline' \h4 measurement before
the SAMI data are matched to LEGA-C, thus enabling us to assess the impact of the
the LEGA-C observing setup on \h4. For these spectra, we use the full wavelength
range of SAMI; note that before measuring the kinematics, we convolve the red arm
to the spectral resolution of the blue arm, using the appropriate Gaussian kernel
\citep{vandesande+2017a}; the gap between the blue- and red-arm spectra is
masked. A comparison between the default \h4 and this baseline value is provided
in \S~\ref{s.r.ss.mbias}.

\subsection{Integrated higher-order kinematics}\label{s.das.ss.hok}

To measure \h4, we follow the procedure outlined in \citetalias{deugenio+2023}. Our measurements
are based on 1-d spectra spanning rest-frame $B$-~and $g$-band; from these data,
we infer the LOSVD using the penalised pixel fitting algorithm
{\sc \href{https://pypi.org/project/ppxf/}{\ppxf}} \citep{cappellari2017,
cappellari2022}.
We configured \ppxf to model the line-of-sight velocity distribution (LOSVD) as
a 4\textsuperscript{th}-order Gauss-Hermite series
\citep{vandermarel+franx1993, gerhard1993}. \ppxf then models the spectra using
a linear combination of simple stellar population (SSP) spectra from the MILES
library \citep{vazdekis+2010, vazdekis+2015}, using BaSTI isochrones
\citep{pietrinferni+2004, pietrinferni+2006} and solar $[\alpha/\mathrm{Fe}]$.
As alternatives to MILES SSP library, we also use the MILES stellar library
\citep{falcon-barroso+2011}, the IndoUS stellar library \citep{valdes+2004}, and
the C3K library \citep{conroy+vandokkum2012} with MIST isochrones
\citep{dotter2016, choi+2016}.
We concluded that the choice of library does not
affect our results (see \S~\ref{s.r.ss.mbias}). However, as discussed in
\citetalias{deugenio+2023}, we adopt the MILES SSP library as default because it provides the
highest fidelity in reproducing the observed spectra \citep[in agreement with
other authors, e.g.][]{vandesande+2017a, maseda+2021}.
In our setup, \ppxf returns the first (non-trivial) four moments of the LOSVD:
mean velocity $V$, velocity dispersion $\sigma$, $h_3$ (a measure of skewness)
and \h4 (measuring excess kurtosis). Note that, throughout this article, we
only use \h4 and ignore the other three measurements, including $\sigma$. In
particular, when we use the second moment \sigap, its values are derived from
other sources, which model the LOSVD as a Gaussian (see \S~\ref{s.das.ss.anc}).

In principle, because the focus of this work are quiescent galaxies, our spectra
should be free from emission lines. In practice, however, even quiescent
galaxies can display strong emission lines, particularly at higher redshift
\citep{maseda+2021}. Moreover, we will consider a subset of higher-redshift
star-forming galaxies as substitutes for the progenitors of local quiescent
galaxies. For these reasons, we mask the wavelength regions where gas emission
lines may arise, regardless of whether any emission was actually detected. This
ensures a homogeneous treatment of all galaxies, but we note that simultaneous
fitting of the emission lines or \textsigma-clipping-based rejection do not change
our conclusions \citepalias{deugenio+2023}.

A key feature of \ppxf is penalisation against non-Gaussian LOSVDs, to recover
reliable solutions in low-$S/N$ data. The amount of penalisation is determined
by the keyword {\sc bias}, which we set to its default value. As explained in
\citetalias{deugenio+2023}, this choice does not affect our measurements of \h4, because of the
high $S/N$ of our spectra; the average $S/N$ of our sample is even higher than
of \citetalias{deugenio+2023}, so the effect of {\sc bias} on the \h4 measurements must be
smaller.

During the data homogenisation process, the seeing matching and slit convolution
introduce correlations between the pixels, but we do not track this in the
noise spectrum. To compensate for correlated noise, we repeat each fit twice.
In the first iteration, we use uniform weights for all (valid) pixels. This step
enables us to measure an empirical $S/N$ by estimating the root mean square of
the residuals in a moving window (see Looser et~al., in~prep.). After this fit,
we rescale the noise spectrum so that the reduced $\chi^2$ is unity. In the
second fit, we use this rescaled noise spectrum and 3-\textsigma clipping to
remove any outliers.
The impact of some of our choices on the results is discussed in
\S~\ref{s.r.ss.mbias}.

As for the measurement uncertainties, we use the default values from \ppxf,
which we checked against Monte-Carlo-derived uncertainties as explained in
\citetalias{deugenio+2023}.

Example \ppxf fits are shown in Fig.~\ref{f.das.ppxf}: in
panels \subref{f.das.ppxf.a} (SAMI 184689), \subref{f.das.ppxf.d} (MAGPI
1206192190) and \subref{f.das.ppxf.g} (LEGA-C 92258~M12). These three galaxies
are randomly selected from the mass-matched sample (defined in
\S~\ref{s.samp.ss.massmat}). In each of the three panels, the grey line is the
galaxy 1-d slit spectrum and the red line is the best-fit \ppxf spectrum, with
\h4 reported in the bottom right corner. The vertical grey lines/regions are
spectral pixels/intervals that have been masked. Panels~\subref{f.das.ppxf.b},
\subref{f.das.ppxf.e} and~\subref{f.das.ppxf.h} show the relative residuals.

\subsubsection{Age-dependent bias}\label{s.das.ss.hok.sss.agebias}

In \citetalias{deugenio+2023}, we have qualitatively shown that \h4 information
is `distributed' in both strong absorption lines as well as less prominent
features. It is a well known fact of stellar evolution that number and prominence
of optical features in an SSP spectrum is a strong function of the SSP age.
This raises the question of how the light-weighted age of a galaxy affects the
fidelity of our \h4 measurements. This question is particularly important for
our work, because we aim at comparing \h4 across different cosmic epochs, when
quiescent galaxies have systematically different stellar population ages.

To understand how stellar population light-weighted age affects our ability to
measure \h4,
we create two sets of mock spectra. These correspond to a quiescent
galaxy with formation redshift $z=3$, observed at $z=0.73$ (the redshift of LEGA-C) and
at $z=0.05$ (the redshift of SAMI). We model the galaxy as a SSP from the
already described MILES library, adopting solar metallicity and either
age equal to 4.5~Gyr (at $z=0.73$) or 10.5~Gyr (at $z=0.05$). For each of these
two spectra, we create two models: one with $\h4 = 0$ and one with $\h4 = 0.06$,
resulting in four model spectra. For each of these four spectra, we create one
thousand random realisations by adding Gaussian noise corresponding to a
$S/N = 20$~\AA$^{-1}$. We then use \ppxf to measure \h4, with the
same setup we used for the real data.

We define $\langle \Delta\,\h4 \rangle$ as the median difference between the
measured and the input value of \h4, and find that, for \h4=0, $\langle
\Delta\,\h4 \rangle = -0.003\pm0.001$. Similarly, for \h4=0.06, we find
$\langle \Delta\,\h4 \rangle = -0.002\pm0.001$. These results apply to both
the 4.5-Gyr-old and the 10.5-Gyr-old mock galaxies. These offsets are
statistically significant, although only to the 3- or 4-\textsigma level. The standard
deviation of the $\Delta\,\h4$ distributions are 0.024 and 0.019 for \h4=0 (for the
4.5-Gyr-old and the 10.5-Gyr-old mocks respectively) and 0.020 and 0.017 for
\h4=0.06 (for the 4.5-Gyr-old and the 10.5-Gyr-old mocks respectively). As
expected from considerations about the depth of absorption features, at fixed
$S/N$, the scatter is larger for the younger population. The difference however
is not dramatic (only $\approx 20$~per\ cent). What is more important, is that in all
four cases, the offset is negligible compared to other sources of systematic
errors (such as stellar template libraries, see \citetalias{deugenio+2023}). Moreover, the
standard deviation of each of the four $\Delta\,\h4$ distributions is 2--3
times smaller than the precision threshold for our quality selection
(\S~\ref{s.samp.ss.qcsel}).

\subsubsection{Measurement bias}\label{s.das.ss.hok.sss.mbias}

\begin{figure}
  \includegraphics[type=pdf,ext=.pdf,read=.pdf,width=1.\columnwidth]{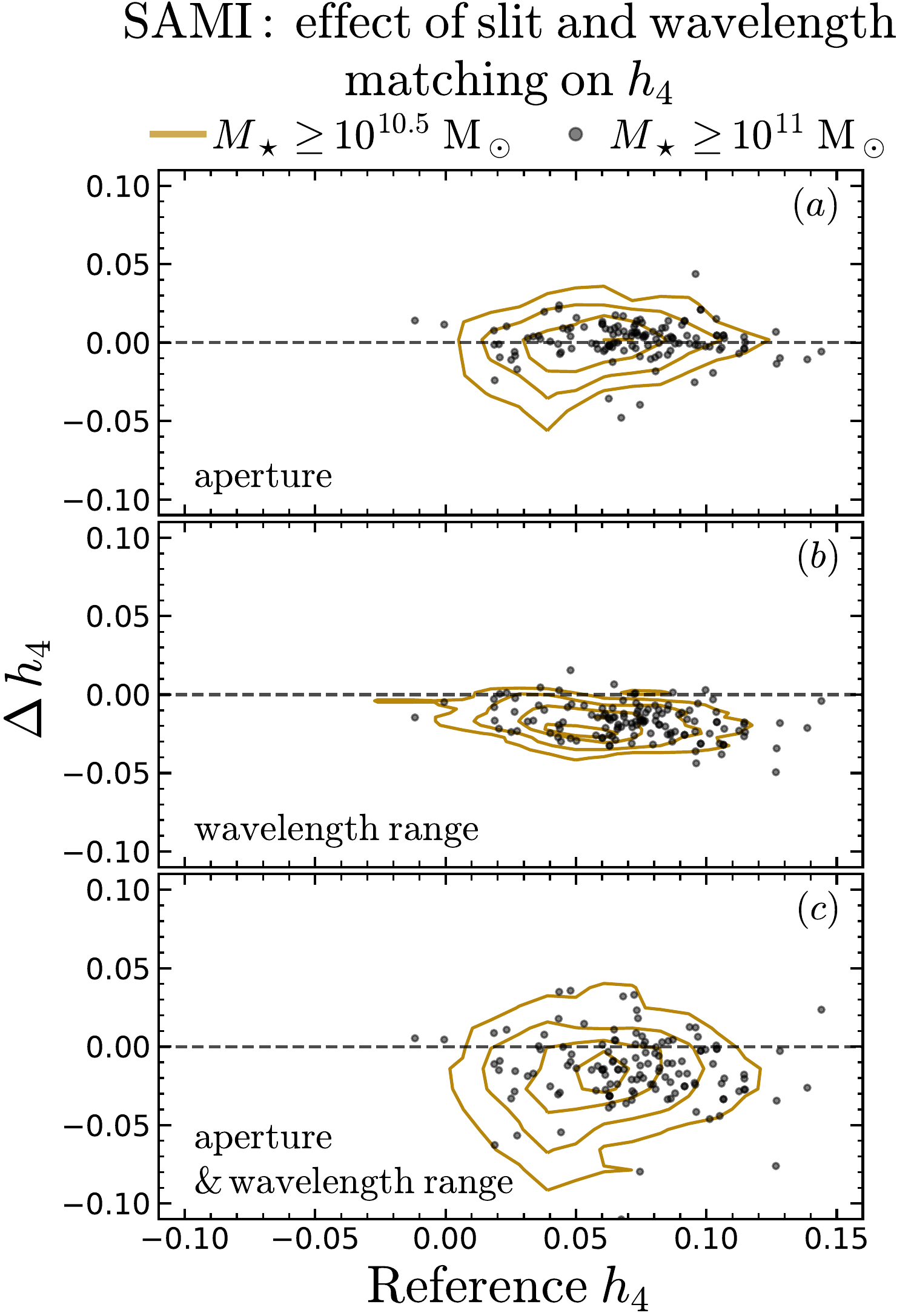}
  {\phantomsubcaption\label{f.das.h4bias.a}
   \phantomsubcaption\label{f.das.h4bias.b}
   \phantomsubcaption\label{f.das.h4bias.c}
  }
  \caption{Effect of different measurement configurations on the reference
  value of \h4 for SAMI quiescent galaxies. The sand contours mark the
  distribution of SAMI quiescent galaxies with $\mstar \geq 10^{10.5} \,
  \msun$ (the contours enclose the 
  30\textsuperscript{th}, 50\textsuperscript{th}, and~90\textsuperscript{th}
  percentiles of the data); the black dots are the mass-matched sample
  ($\mstar \geq 10^{11} \, \msun$). The reference value is
  measured inside the elliptical aperture of semi-major axis equal to
  one~\re, using the MILES SSP library and the full wavelength range of 
  SAMI. Panel~\subref{f.das.h4bias.a} shows the effect of using a slit instead
  of the elliptical aperture (including seeing convolution, see
  \S~\ref{s.das.ss.match}); panel~\subref{f.das.h4bias.b} shows the effect of
  restricting the wavelength range to the wavelength range of LEGA-C; and
  panel~\subref{f.das.h4bias.c} combines both the slit setup and restricted
  wavelength range, thus matching our default
  measurements. This figure underscores the importance of our data
  homogenisation (\S~\ref{s.das.ss.match}).
  }\label{f.das.h4bias}
\end{figure}

Because we aim to compare measurements between different cosmic epochs and
different surveys, we need to understand how different data quality and
instrument setup affect the value of \h4. To address this question, we leverage
IFU spectroscopy from SAMI. For each galaxy, we
define a reference \h4 measurement from the aperture spectrum inside the
elliptical aperture of semi-major axis equal to one~\re. These spectra cover
the whole wavelength range of SAMI. We then compare this reference value to
the default measurement, obtained from 1-d synthetic-slit spectroscopy and
designed to match the observing setup of LEGA-C (as described in
\S~\ref{s.das.ss.match}). We split the data homogenisation procedure in two 
steps: aperture matching and wavelength matching. In Fig.~\ref{f.das.h4bias.a},
we compare the reference \h4 values to the \h4 we measure from the slit
setup (including seeing convolution); the golden contours trace the
distribution of SAMI quiescent galaxies with $\mstar \geq 10^{10.5} \,
\mathrm{M}_\odot$, the dots are quiescent SAMI galaxies from the mass-matched
sample ($\mstar \geq 10^{11} \, \mathrm{M}_\odot$). We find that, for both
stellar-mass selections, the slit setup does not bias \h4; considering only
massive galaxies (black dots in Fig.~\ref{f.das.h4bias}), the median
difference between this \h4 and the reference value is  $\langle \Delta
\, \h4 \rangle = 0.003 \pm 0.001$, with a root mean square (rms) of 0.010.
In panel~\subref{f.das.h4bias.b}, we compare the reference \h4 to the value we
measure from the same aperture, but using only the bluest wavelengths, to
match the rest-frame range of LEGA-C. In this case, the picture is opposite
to what we found in panel~\subref{f.das.h4bias.a}: the offset is large but the
scatter is small: the median difference is $\langle \Delta \, \h4 \rangle =
-0.017 \pm 0.001$ and the rms is 0.011.
Finally, in panel~\subref{f.das.h4bias.c}, we combine both slit aperture and
wavelength matching, thus obtaining our homogenised, default \h4 values. As
expected, we find both a systematic offset ($\langle \Delta \, \h4 \rangle =
-0.015 \pm 0.002$) and a large rms (0.019).

\subsection{Stellar masses and ancillary data}\label{s.das.ss.anc} 

The ancillary data we use in this work are the same as in \citetalias{deugenio+2023}, to which
we refer for a more detailed discussion. Here we provide a summary and the
most important remarks.

For SAMI, stellar masses \mstar are derived from the $i$-band total
magnitude and a mass-to-light ratio based on $g-i$ colour \citep{taylor+2011}.
Star-formation rates (SFR) are derived from the attenuation-corrected H\textalpha\
luminosity \citep{croom+2021a}.
For MAGPI, we use \mstar and SFR full SED fits from {\sc prospect}
\citep{robotham+2020}.
We then define SAMI and MAGPI quiescent galaxies as lying more than 1~dex below
the star-forming sequence of \citet{whitaker+2012}.
For LEGA-C, we use SED fits to observed-frame $BVrizYJ$ photometry, based on
the {\sc prospector} software \citep{leja+2019a, johnson+2021}, as explained
in \citet{vanderwel+2021}. We compared these measurements using a common subset
with \mstar derived from {\sc magphys}, and found systematic differences of
$0.03$~dex with a scatter of $0.07$~dex, which are acceptable for the goals of
this article \citepalias{deugenio+2023}.
 
In this article we consider primarily quiescent galaxies (but the
progenitor-matched sample from LEGA-C also includes star-forming galaxies,
see \S~\ref{s.samp.ss.progmat}). Quiescent galaxies are defined as lying
1.6~dex below the star-forming sequence (for SAMI and MAGPI galaxies with
$z<0.41$), as having H\textbeta\ equivalent width $>-1$~\AA\ (for MAGPI galaxies
with $z>0.41$, where the MUSE spectral range does not cover H$\alpha$), or
as lying in the quiescent corner of the UVJ diagram (for LEGA-C). This
classification follows \citetalias{deugenio+2023}.

Galaxy sizes and shapes are derived from the best-fit S{\'e}rsic models to
observed frame $r-$band photometry (for SAMI and MAGPI) and to HST ACS F814W
photometry (for LEGA-C). In particular, galaxy sizes
are the half-light semi-major axis of the model. We note that for SAMI,
replacing $r-$band photometry with $g-$band photometry does not change our
results \citepalias{deugenio+2023}.

In addition to our main selection based on \mstar, we use two alternative
methods based on aperture velocity dispersion \sigap, virial mass
$M_\mathrm{vir}$ and total masses from dynamical models $M_\mathrm{JAM}$.
For SAMI and LEGA-C, \sigap is taken from the literature.
For SAMI, it is the value measured inside one effective radius \citep{
croom+2021a}. For LEGA-C, it is the value measured from the 1-d slit spectra
\citep{bezanson+2018a, vanderwel+2021}. For MAGPI, we use our own
measurements obtained with \ppxf, using the IndoUS stellar library and
assuming a Gaussian LOSVD, for consistency with the \sigap measurements of
SAMI and LEGA-C.
For $M_\mathrm{vir}$, we use the expression of \citet{vanderwel+2022},
which uses semi-major axis half-light sizes, incorporates a correction for
non-homology \citep[based on the S{\'e}rsic index $n$][]{cappellari+2006}
and adds an inclination correction.
The dynamical models are based on Jeans Anisotropic Models
\citep[JAM,][]{cappellari2008}, adapted to slit spectroscopy as described in
\citet{vanhoudt+2021}. Note that $M_\mathrm{JAM}$ is only available for
roughly one third of the sample, because galaxies where the photometric
major axis is not aligned to the slit were not modelled.

\section{Sample selection}\label{s.samp}

In this section, we aim to present the motivation, selection criteria and
properties of the two samples we use in this work. The first sample, consisting
of massive, quiescent galaxies from SAMI, MAGPI and LEGA-C, is the `mass-matched
sample', aiming at characterising the evolution of the quiescent galaxy
population over cosmic time, including both demographic changes and physical
evolution. The second sample is the `progenitor-matched sample', selected
from SAMI and LEGA-C to provide a plausible connection between local quiescent
galaxies and their progenitors.

We start by illustrating the quality selection criteria
(\S~\ref{s.samp.ss.qcsel}), then we describe the `mass-matched sample'
(\S~\ref{s.samp.ss.massmat}) and the `progenitor-matched sample'
(\S~\ref{s.samp.ss.progmat}).

\subsection{Quality selection}\label{s.samp.ss.qcsel}

Because in this work we focus on massive, quiescent galaxies, we consider only
galaxies with $\mstar > 10^{11} \, \msun$ \citep[which is also the
completeness limit of LEGA-C][]{vanderwel+2021}. Above this threshold, we have
211, 22 and 1027 galaxies from SAMI, MAGPI and LEGA-C
\citep[for LEGA-C, we consider only primary galaxies with flag {\sc use}=1][]{
straatman+2018}.

Following \citetalias{deugenio+2023}, we impose a quality cut at $u(\h4)<0.05$, to ensure a
reliable measurement of \h4. After this cut, we have 200, 22 and 692 galaxies
for the three surveys; from these, we consider only quiescent galaxies, arriving
to a final sample of 135, 22 and 479 targets for SAMI, MAGPI and LEGA-C,
respectively. The completeness of each survey relative to the samples before
any quality cut is 99, 100 and 90~per\ cent.

For the progenitor-matched sample, we consider LEGA-C galaxies with
$\mstar > 10^{10.5} \, \mathrm{M}_\odot$, which is below the mass cut of the
main sample at  $\mstar > 10^{11} \, \mathrm{M}_\odot$. This cut is chosen to account
for the increase in stellar mass between the epochs of LEGA-C and SAMI. The
progenitor pool consists of both star-forming and quiescent galaxies, with lower
completeness compared to the main sample. In particular, after imposing $u(\h4)<0.05$, the
completeness for LEGA-C quiescent galaxies is 81~per\ cent, and for star-forming
galaxies is only 28~per\ cent \citetext{see \citetalias{deugenio+2023}, their fig.~4}. Completeness
decreases with decreasing \mstar, which means our results for the
progenitor-matched sample are likely a conservative estimate (see
\S~\ref{s.r.ss.sbias}).

\subsection{The mass-matched sample from SAMI, MAGPI and LEGA-C}\label{s.samp.ss.massmat}

\begin{figure}
  \includegraphics[type=pdf,ext=.pdf,read=.pdf,width=1.\columnwidth]{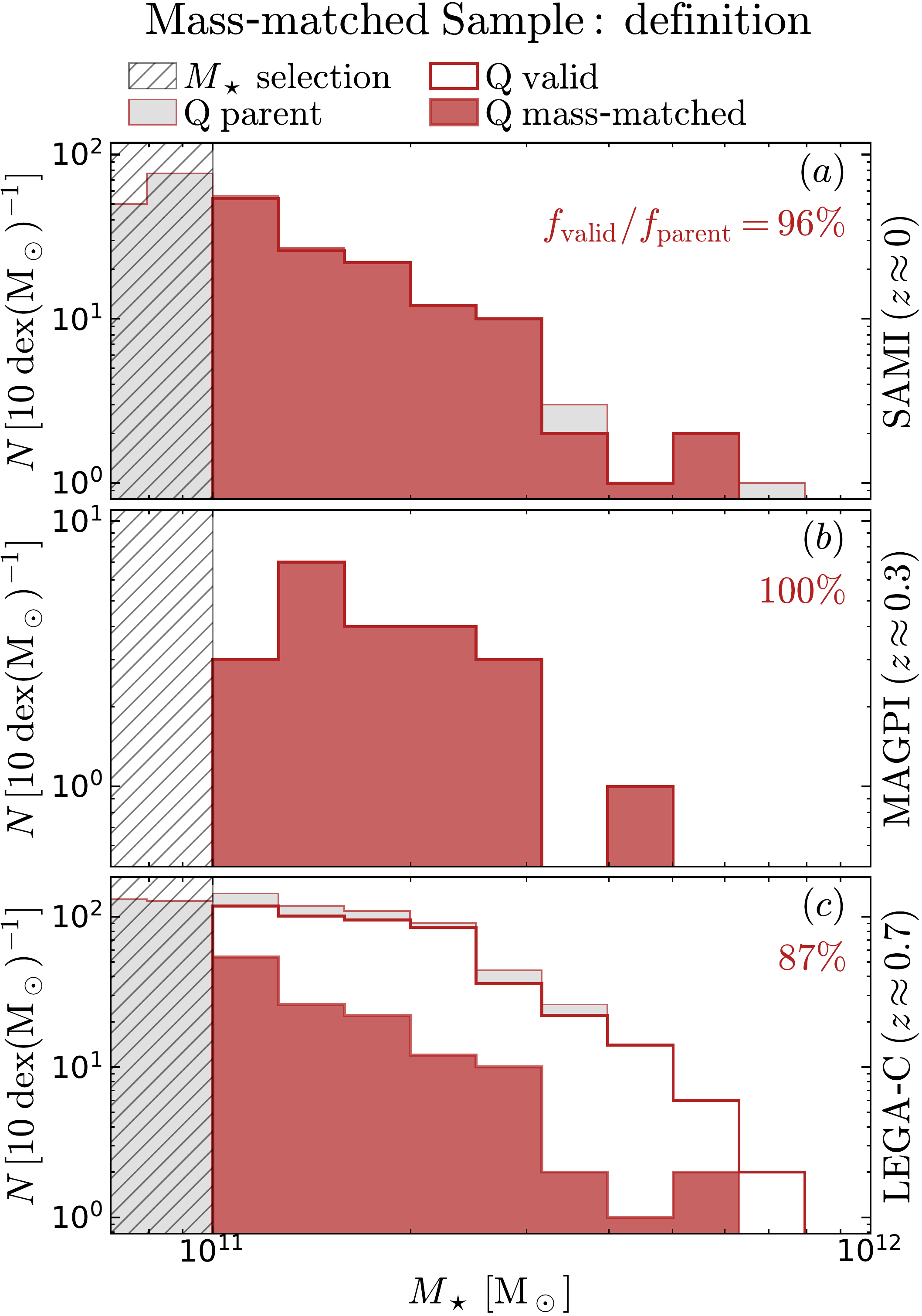}
 {\phantomsubcaption\label{f.samp.massmat.a}
  \phantomsubcaption\label{f.samp.massmat.b}
  \phantomsubcaption\label{f.samp.massmat.c}}
  \caption{Stellar mass distribution of the mass-matched sample. The three
  rows show the SAMI, MAGPI and LEGA-C sample. The dashed regions show the
  cut at $\mstar > 10^{11} \, \mathrm{M}_\odot$, the filled grey histograms
  are the parent samples, consisting of all quiescent (Q) galaxies above the
  mass cut. The solid red empty histograms are the valid samples, consisting
  of all
  quiescent galaxies above the mass and quality selection cuts. The filled
  red histogram is the mass-matched sample; for SAMI and MAGPI, this coincides
  with the valid sample; for LEGA-C, it consists of a subset of the valid
  sample, chosen to reproduce the mass distribution of the SAMI valid sample.
  }\label{f.samp.massmat}
\end{figure}

The first task we aim to accomplish is to compare the \h4 distribution of
galaxies at fixed \mstar. This defines the `mass-matched sample'
(Fig.~\ref{f.samp.massmat}). Physically, a selection at fixed \mstar is a
logical contradiction, because the stellar mass of central galaxies increased
with cosmic time. However, this sample provides a
conservative, `minimum-baseline' measurement for any evolution of \h4, free
from the assumptions needed to connect local galaxies with
progenitor-like galaxies at higher redshift (which we do in
\S~\ref{s.samp.ss.progmat}). In particular, the mass-matched sample puts together
local galaxies with over-massive (false) progenitors at higher redshift; this fact,
together with the \h4--\mstar correlation \citepalias{deugenio+2023}, means that the mass-matched
sample is biased to find \h4 decreasing with cosmic time.

The three samples are matched in \mstar (as closely as
possible). 
We match the LEGA-C sample to the mass function of SAMI, because,
among these two surveys, it is LEGA-C that has the largest volume. For MAGPI,
where the effective survey volume is relatively small, we resort to taking all
galaxies in the same mass range as SAMI, without matching the mass function.

For SAMI, the parent sample consists of 141 quiescent galaxies with 
$\mstar \geq 10^{11} \, \msun$ (filled grey histogram in
panel~\subref{f.samp.massmat.a}). From this sample, we remove six with $u(\h4)
\geq 0.05$, arriving to a valid sample of 135 galaxies. This is the valid
sample, which for SAMI coincides with the mass-matched sample (so in 
panel~\subref{f.samp.massmat.a} the solid red empty and solid red filled
histograms coincide).

For MAGPI, we have 22 massive, quiescent galaxies, all of which meet the
quality selection. Given this small sample size, we do not resample this set to
match the mass distribution of SAMI (so, for MAGPI too, the valid and
mass-matched samples coincide, panel~\subref{f.samp.massmat.b}).

For LEGA-C, the parent sample contains 530 quiescent galaxies with $\mstar \geq
10^{11} \, \msun$, of which 479 meet the quality selection
criterion (solid red empty histogram in panel~\subref{f.samp.massmat.c}). We
then weight these 479 galaxies to match the mass function of SAMI; the weights
are simply the ratio between the SAMI and LEGA-C valid histograms.
The resulting (weighted) mass distribution of the resulting sample is traced by
the solid red filled histogram in panel~\subref{f.samp.massmat.c}.
Two galaxies outside the mass range of SAMI get weight zero, so they are
effectively removed from the mass-matched LEGA-C sample. For the others, we
rescale the weights so that they add to the sample size of LEGA-C (this preserves
the relative weight of different samples when we combine them). After this
rescaling, the minimum and maximum weights are 0.25 and 2.36.

\subsection{The progenitor-matched sample from SAMI and LEGA-C}\label{s.samp.ss.progmat}

\begin{figure}
  \includegraphics[type=pdf,ext=.pdf,read=.pdf,width=1.\columnwidth]{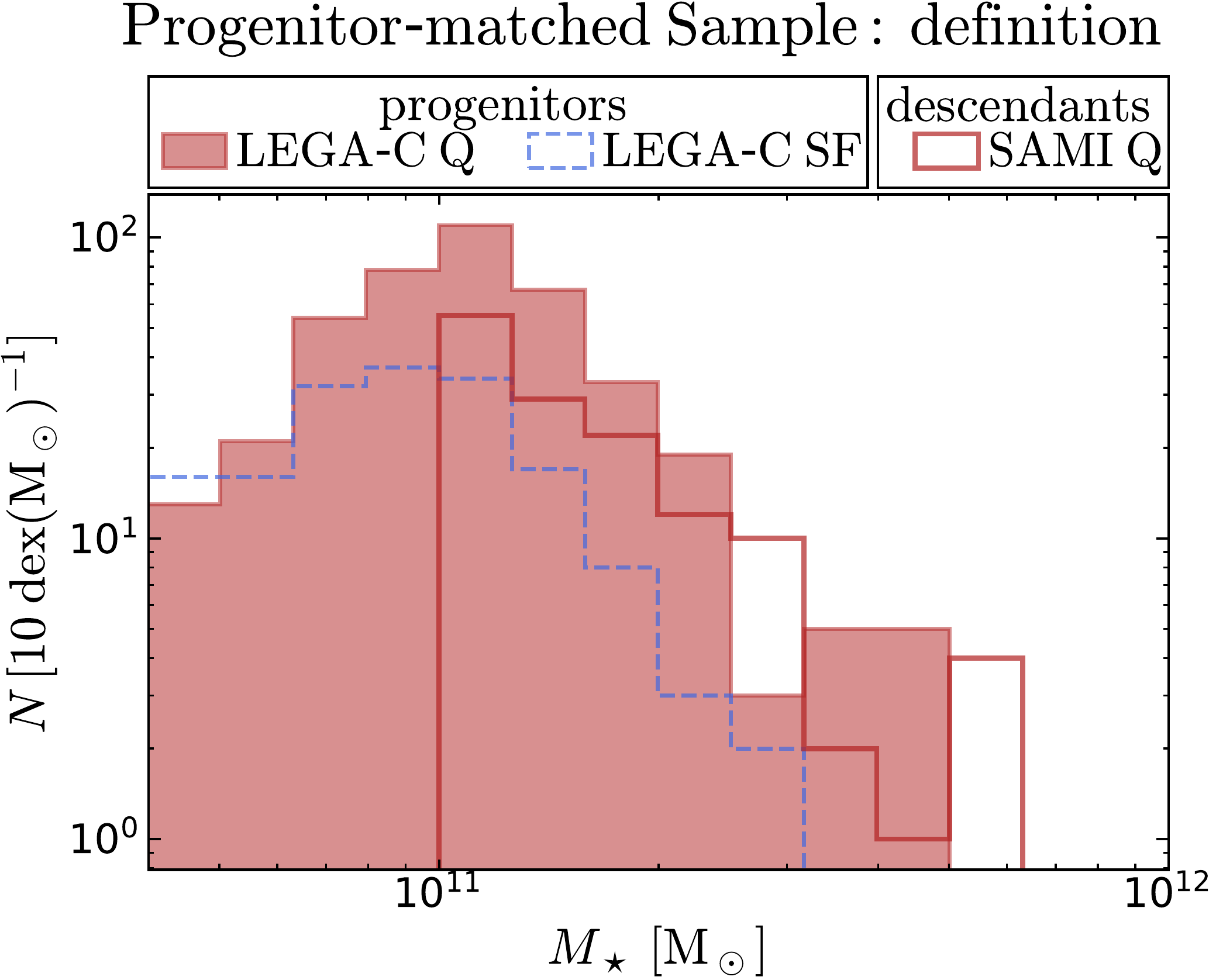}
  \caption{The `progenitor-matched' sample consists of local massive quiescent
  galaxies (from SAMI, solid red empty histogram) and their possible $z=0.7$
  progenitors drawn from LEGA-C, including both star-forming galaxies
  (dashed blue histogram) and quiescent galaxies (red filled histogram). The
  mass distribution of the progenitors was derived from IllustrisTNG, but we
  note that, in the simulation, the fraction of quiescent progenitors is
  only 30~per\ cent.
  }\label{f.samp.progmat}
\end{figure}

While the mass-matched sample addresses the issue of mass-related bias, we know
that galaxies evolve in both mass and size, even after becoming quiescent
\citep[e.g.][]{taylor+2010, vanderwel+2014}. To address the effect of this
evolution, we use data from the publicly available
IllustrisTNG simulations to inform the mass distribution of the progenitors of
the SAMI sample \citetext{\citealp{marinacci+2018}, \citealp{naiman+2018},
\citealp{nelson+2018}, \citealp{springel+2018} and \citealp{pillepich+2018}}.
We then use the progenitors' mass distribution to select progenitor-like galaxies
from LEGA-C. We do not attempt any match for MAGPI, so this survey is not part
of the `progenitor-matched sample'.

Following \citet{rodriguez-gomez+2015}, we define a galaxy as a subset of the
simulation volume identified by the {\sc subfind} algorithm
\citep{springel+2001, dolag+2009} and use the combined mass of all stellar
particles as the galaxy stellar mass.
While this is not immediately comparable to observed stellar masses,
even truncating at a galaxy-dependent aperture may introduce unwanted bias
\citep{degraaff+2022}. The effect of this aperture bias is secondary compared to
the inclusion of star-forming galaxies in the progenitor sample.

We consider data from the run TNG100-1, then take all quiescent galaxies from
snapshot 94 ($z=0.06$, matching SAMI) and trace their `main-branch' progenitors
to snapshot 58 \citep[$z=0.73$, matching LEGA-C. Main-branch progenitors are
defined as the progenitors with the most massive history behind them;][]{
delucia+blaizot2007, rodriguez-gomez+2015}. We then randomly select quiescent
galaxies from snapshot 94 matching the mass distribution of the SAMI sample (i.e., following
the solid red empty histogram in Fig.~\ref{f.samp.progmat}). This matching
procedure is repeated 100 times to explore all realisations of the random
sampling.
We then obtain the
mass distribution of their snapshot-58 progenitors. Of these, most
($>$95~per\ cent) have $\mstar > 10^{10.5} \, \mathrm{M}_\odot$. Among all
random realisations, on average only 33~per\ cent of progenitors are already
quiescent at $z=0.73$, in agreement with
the results of \citet[][cf. their fig.~16]{moster+2020}.
We then sample the LEGA-C galaxies to match the mass distribution of the
IllustrisTNG `progenitors' of the SAMI quiescent galaxies. This is done separately
for star-forming and quiescent galaxies and considering only LEGA-C galaxies above
the adopted quality cuts (\S~\ref{s.samp.ss.qcsel}).
The mass distribution of the LEGA-C `progenitors' of SAMI quiescent galaxies is
shown in Fig.~\ref{f.samp.progmat}, with the dashed blue and red filled histogram
representing the star-forming and quiescent subsets.
It is clear that our sample of progenitors has the wrong ratio between star-forming
and quiescent galaxies: we select 170 and 410 galaxies from each class, respectively,
whereas only 33~per cent of the simulated progenitors was already quiescent at
$z=0.73$. This discrepancy is dictated by the availability of LEGA-C galaxies.
To correct for it, we apply weights in bins of \mstar whenever calculating the
median \h4 of the progenitor sample, or the Pearson correlation coefficient for the
evolution of \h4 (\S~\ref{s.r.ss.h4z}).

A crucial caveat of the progenitor-matched sample is that selecting progenitors based
solely on stellar mass does not completely remove progenitor bias. To prove this, we
randomly select $z=0.73$ IllustrisTNG galaxies with the same criterion used to select
the LEGA-C progenitor-matched sample; i.e., we select galaxies from snapshot 54
following the same distribution of stellar mass as the progenitors of
the $z=0$ IllustrisTNG galaxies (which were in turn selected to match the SAMI
stellar mass distribution).
Looking at the descendants of these simulated galaxies, we find the following.
Between snapshots 58 and 94, the quiescent progenitors remain mostly quiescent
(16~per cent rejuvenate). Most (69~per cent) of the star-forming progenitors
become quiescent. The fraction of progenitors which underwent at least one major
merger between snapshots 58 and 94 is 24~per cent (mass ratio greater or equal to
1/3; using mass ratios of 1/2 and 2/3 the fractions are 17 and 12~per cent).

\section{Cosmic evolution of \texorpdfstring{\h4}{h4}}\label{s.r.ss.h4z}

\begin{figure*}
  \includegraphics[type=pdf,ext=.pdf,read=.pdf,width=1.\textwidth]{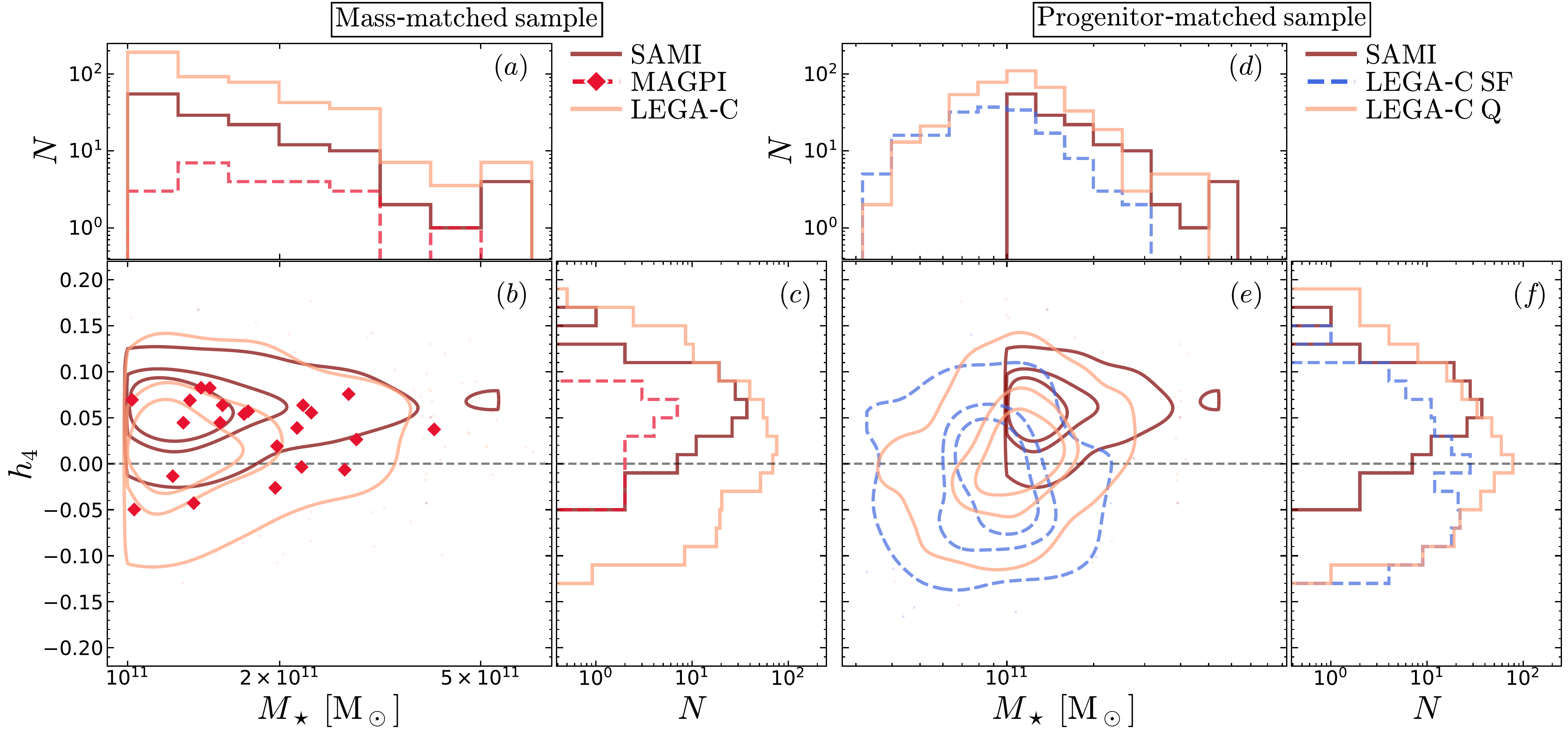}
  {\phantomsubcaption\label{f.r.h4m.a}
   \phantomsubcaption\label{f.r.h4m.b}
   \phantomsubcaption\label{f.r.h4m.c}
   \phantomsubcaption\label{f.r.h4m.d}
   \phantomsubcaption\label{f.r.h4m.e}
   \phantomsubcaption\label{f.r.h4m.f}}
  \caption{Relation between \h4 and \mstar for the mass-matched sample
  (panels~\subref{f.r.h4m.a}--\subref{f.r.h4m.c}) and the progenitor-matched
  sample (panels~\subref{f.r.h4m.d}--\subref{f.r.h4m.f}). In
  panels~\subref{f.r.h4m.b} and~\subref{f.r.h4m.e}, the contour lines
  represent the 30\textsuperscript{th}, 50\textsuperscript{th} and
  90\textsuperscript{th} percentiles of the data (MAGPI galaxies are
  represented individually as diamonds).
  Despite having the same \mstar distribution (panel~\subref{f.r.h4m.a}), the
  SAMI and LEGA-C samples have different \h4 distributions
  (panel~\subref{f.r.h4m.c}). The difference in \h4 is even larger if we
  compare the SAMI sample to the progenitor-matched samples, where the
  different \mstar distribution (panel~\subref{f.r.h4m.d}) amplify the
  difference in \h4.
  }\label{f.r.h4m}
\end{figure*}

In Fig.~\ref{f.r.h4m} we show the relation of \h4 with \mstar
for the mass-matched sample (panels~\subref{f.r.h4m.a}--\subref{f.r.h4m.c}) and
the progenitor-matched sample (panels~\subref{f.r.h4m.d}--\subref{f.r.h4m.f}).
Panel~\subref{f.r.h4m.a} shows the mass distribution of the three samples; by
construction, the SAMI and LEGA-C samples have the same distribution.
In panel~\subref{f.r.h4m.b} we show the relation between \mstar and \h4: there
is little to no evidence for a correlation, at variance with what reported by
\citetalias{deugenio+2023}. This disagreement is due to the different \mstar selection: we
consider only $\mstar > 10^{11} \, \msun$, so we we have a shorter baseline in
\mstar compared to \citetalias{deugenio+2023}. Despite matching in \mstar, the SAMI and
LEGA-C samples have different \h4 distributions (panel~\subref{f.r.h4m.c}).

For the progenitor-matched sample, the difference is even larger (cf.\
panels~\subref{f.r.h4m.c}--\subref{f.r.h4m.f}), because the difference already
reported at fixed \mstar is amplified by the combination between the
\h4--\mstar correlation, and the fact that progenitor-like galaxies have
necessarily lower \mstar than their descendants.
Note that \h4 differs even between star-forming and quiescent progenitors, in
agreement with \citetalias{deugenio+2023}.

\begin{figure*}
  \includegraphics[type=pdf,ext=.pdf,read=.pdf,width=1.\textwidth]{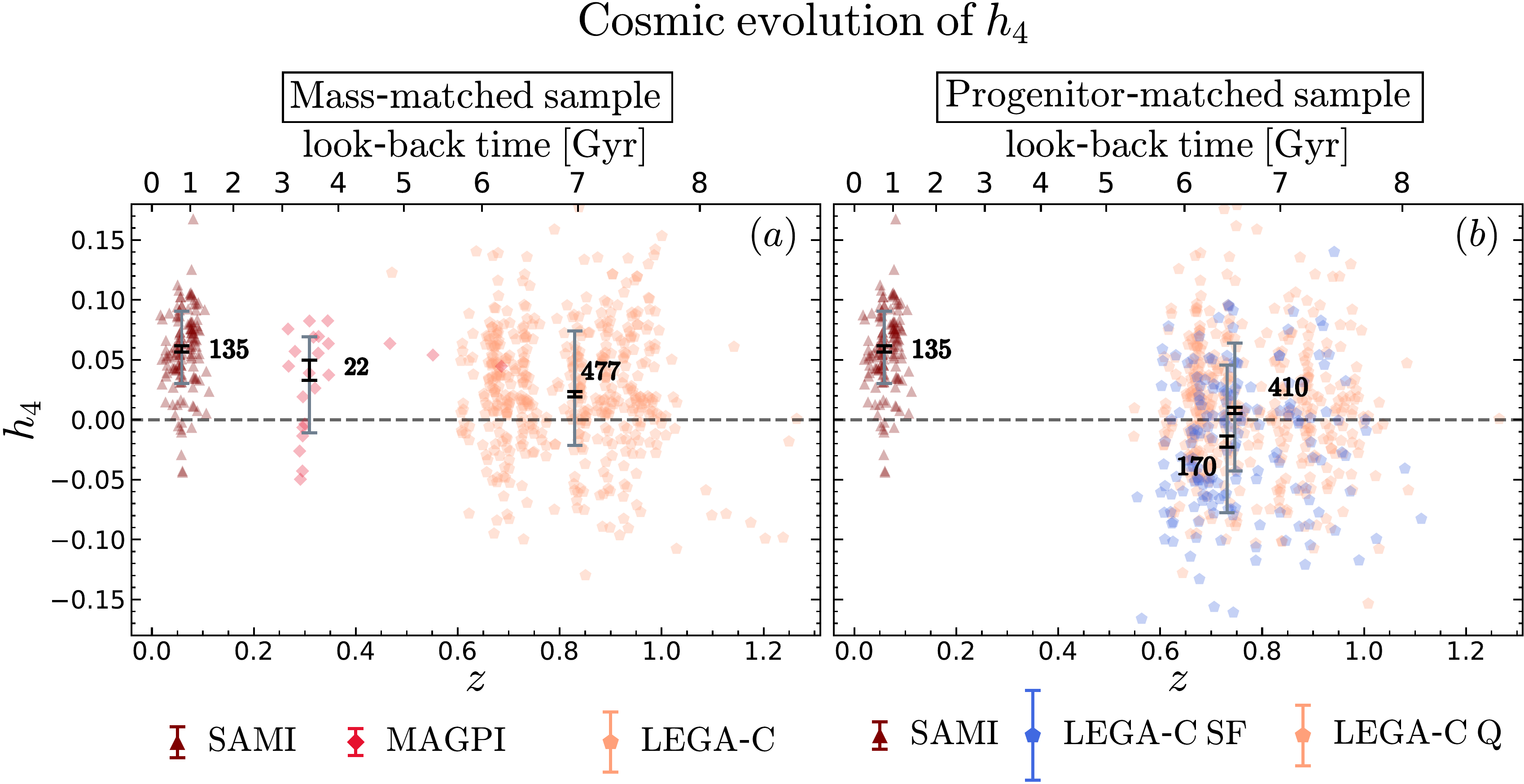}
  {\phantomsubcaption\label{f.r.h4evo.a}
   \phantomsubcaption\label{f.r.h4evo.b}
  }
  \caption{Cosmic evolution of the integrated \h4 moment for massive,
  quiescent galaxies ($\mstar > 10^{11} \, \msun$). Panel~\subref{f.r.h4evo.a}
  shows SAMI (dark red triangles) compared to MAGPI (red diamonds) and to the
  mass-matched sample from LEGA-C (pink pentagons); median measurement
  uncertainties are shown in the legend. The dashed horizontal line corresponds
  to a Gaussian line-of-sight velocity distribution. The grey errorbars
  encompass the 16\textsuperscript{th}--84\textsuperscript{th} percentiles of
  each sample and are located at the median redshift of the sample (with a small
  offset for readability). The black errorbars represent the uncertainty about
  the median value of each sample. The weighted Pearson correlation coefficient
  between all points is $\rho = -0.29$, with a significance of over 7~\textsigma.
  In panel~\subref{f.r.h4evo.b}, SAMI (dark red triangles) is compared to
  progenitor-like galaxies selected from LEGA-C, including both quiescent (pink
  pentagons) and star-forming galaxies (blue pentagons). Accounting for
  progenitor bias, the evolution in \h4 is even stronger
  (to calculate $\rho$, low-mass star-forming galaxies in LEGA-C are upweighted
  to match the mass distribution of the SAMI progenitors, see \S~\ref{s.samp.ss.progmat}).
  }\label{f.r.h4evo}
\end{figure*}

We now consider the redshift evolution of \h4. Fig.~\ref{f.r.h4evo.a} shows \h4
as a function of redshift for the mass-matched sample (\S~\ref{s.samp.ss.massmat}).
SAMI, MAGPI and LEGA-C
galaxies are represented respectively by dark red triangles, red diamonds and
pink pentagons. For each of the three surveys, the median uncertainties on \h4
are represented by the errorbars with the same symbol and colour as the relevant
survey. The three grey errorbars mark the median $z$ and \h4 of each of the
three subsamples; the smallest errorbars are the uncertainties about the median,
the largest errorbars are the 16\textsuperscript{th}--84\textsuperscript{th}
percentiles of the \h4 distribution. Considering the population of massive,
quiescent galaxies, the median \h4 rises from $0.019\pm0.002$ at $z=0.82$
(LEGA-C), to $0.045\pm0.008$ at $z=0.31$ (MAGPI) to $0.059\pm0.004$ at
$z=0.06$ (SAMI; the uncertainties have been estimated by bootstrapping each
sample one thousand times).
It is clear that the difference between SAMI and LEGA-C
is statistically significant, being almost nine standard deviations \textsigma
away from zero. For MAGPI, the difference from SAMI is not significant
(one~\textsigma), but the difference from LEGA-C is marginally significant
(three~\textsigma). The median \h4 for MAGPI is intermediate between SAMI and
LEGA-C, which adds more confidence to the hypothesis that \h4
increases with cosmic time. We do not model the intrinsic scatter of the
distributions, but a simple estimate (by subtracting in quadrature the median
uncertainty from the observed rms) gives an intrinsic scatter of $0.03$ for SAMI
and $0.05$ for LEGA-C.

To establish if there is any evolution, we use the weighted Pearson correlation
coefficient $\rho$ between $z$ and \h4. For SAMI and MAGPI, all weights are set
to one. For LEGA-C, they reflect the relative importance of galaxies in
different bins (\S~\ref{s.samp}).
Considering all three surveys, we find $\rho = -0.29$ ($P=1.2 \times 10^{-13}$,
seven-\textsigma significance).
Removing MAGPI, we get $\rho = -0.30$ and a slightly higher significance ($P=4.2
\times 10^{-14}$).
Further removing SAMI we find $\rho = -0.07$ and $P=0.12$, which is not
significant. In fact, taken separately, none of the three subsets gives a
statistically significant correlation.
Even though we focus on high-mass galaxies, there is nothing special about the
\mstar cut at $10^{11} \, \msun$: if we set the cut at $10^{10.5} \,
\msun$ \citep[i.e. near the completeness limit of LEGA-C,][]{
vanderwel+2021}, we find $\rho = -0.16$ and $P=1.2 \times 10^{-6}$, which is
still statistically significant (4.7~\textsigma).

So far, these results do not account for mass growth between the look-back times
of LEGA-C and SAMI. As we will see, doing so entails including galaxies with
$\mstar < 10^{11}\,\msun$ from the LEGA-C sample, but these galaxies
have even lower average \h4 \citepalias{deugenio+2023}, which makes our result stronger.

The mass-matched analysis can only show the evolution of the quiescent
population as a whole, without taking into account the effect of progenitor bias.
But when we consider the progenitor-matched sample (Fig.~\ref{f.r.h4evo.b}), the
evidence for \h4 increasing with redshift is even stronger. Here the dark red
triangles are the same SAMI galaxies as in panel~\subref{f.r.h4evo.a}, but
blue/pink pentagons are star-forming/quiescent LEGA-C galaxies, chosen to match
the mass distribution of the progenitors of SAMI galaxies (as inferred from
numerical simulations, see \S~\ref{s.samp.ss.progmat}). The meaning of the
errorbars is the same as in panel~\subref{f.r.h4evo.a}. The median \h4 of
quiescent progenitors is $0.011\pm0.003$, and for star-forming progenitors
is $-0.010\pm0.006$; both values are more than nine \textsigma away from the
SAMI median value. These values are different (3-\textsigma significance), in
agreement with the difference in \h4 between star-forming and quiescent galaxies
reported in \citetalias{deugenio+2023}. To combine the quiescent and star-forming progenitors,
we upweight the latter, to account for the fact that two thirds of the simulated
progenitors are star forming, but only one-third of LEGA-C progenitors are
star forming. The weighted median is $-0.006\pm0.004$ which is, as expected,
lower than the value inferred from the mass-matched sample.
Considering all galaxies in the progenitor-matched sample, we find
$\rho = -0.36$, $P=3.5 \times 10^{-23}$. Considering only SAMI and quiescent
progenitors, we find  $\rho = -0.38$, $P=2.6 \times 10^{-20}$, while considering
only SAMI and star-forming progenitors we have $\rho = -0.59$, $P=7.1 \times
10^{-30}$. Overall, the progenitor-matched analysis suggests that when
accounting for progenitor bias, the evolution of \h4 is higher and more
statistically significant.

\subsection{Sample selection bias}\label{s.r.ss.sbias}

\begin{figure*}
  \includegraphics[type=pdf,ext=.pdf,read=.pdf,width=1.\textwidth]{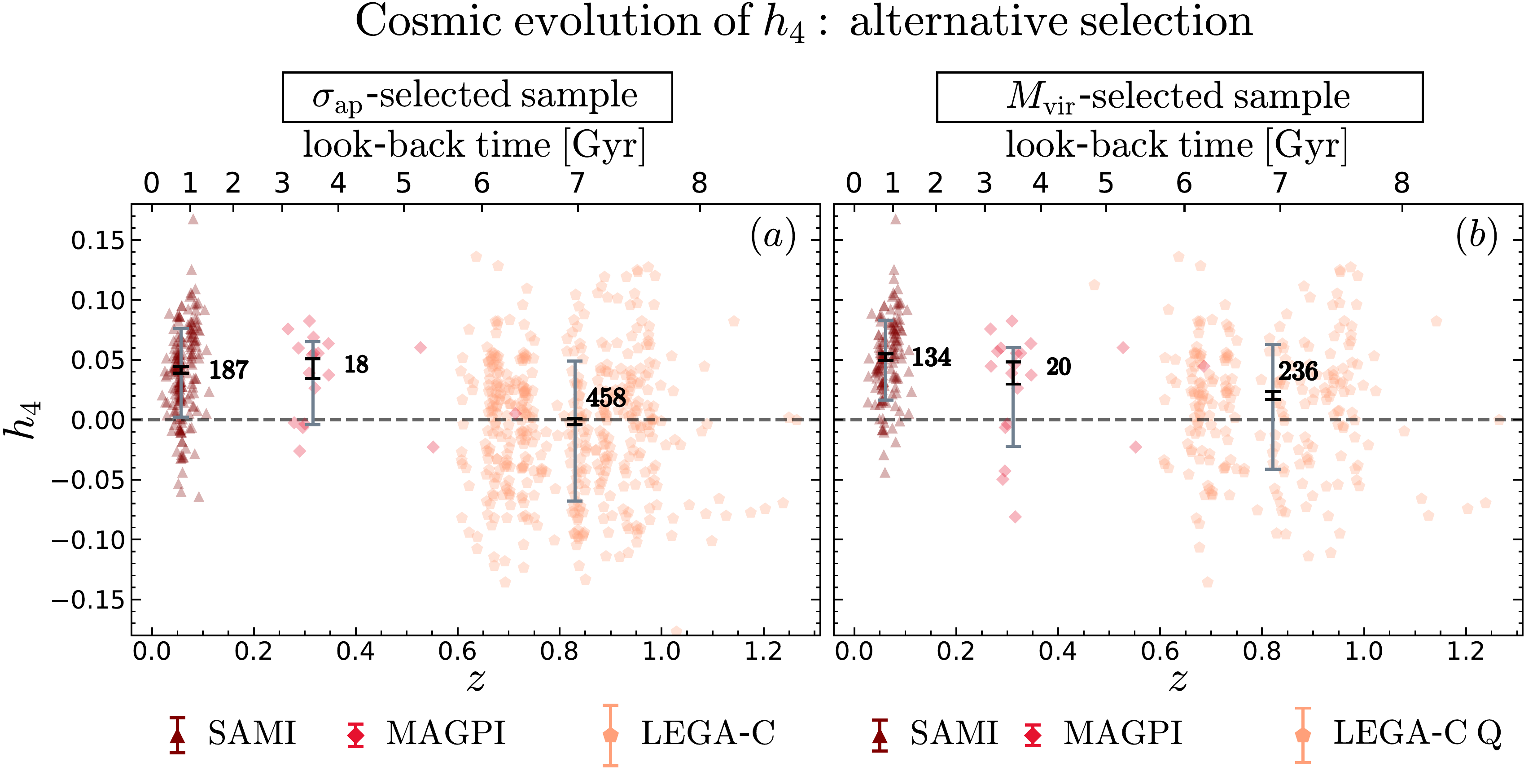}
  {\phantomsubcaption\label{f.r.h4evoalt.a}
   \phantomsubcaption\label{f.r.h4evoalt.b}
  }
  \caption{The finding that \h4 changes with cosmic time is independent
  from our \mstar selection: the trend is observed even if we select quiescent
  galaxies based on \sigap (panel~\subref{f.r.h4evoalt.a}) or $M_\mathrm{vir}$
  (panel~\subref{f.r.h4evoalt.b}). The meaning of the symbols is the same
  as Fig.~\ref{f.r.h4evo.a}. Note here we did not match the sample in
  their \sigap distribution (for panel~\subref{f.r.h4evoalt.a}) or
  $M_\mathrm{vir}$ distribution (for panel~\subref{f.r.h4evoalt.b}).
  }
\end{figure*}\label{f.r.h4evoalt}

Could systematic errors in the measurement of \mstar explain the trend between
\h4 and redshift? Given the correlation between \h4 and \mstar \citepalias{deugenio+2023},
and given the heterogeneous photometric data available to measure \mstar, this
is a reasonable concern. We adopt a conservative approach,
by repeating the mass-matched analysis with a cut $\mstar > 10^{10.5} \,
\msun$ \emph{for SAMI only}. This cut represents the extreme
hypothesis of \mstar being systematically underestimated by a factor of 3 at
$z=0$ compared to the redshift $z=0.8$ of LEGA-C. As expected from the
\h4--\mstar correlation, the average \h4 for SAMI decreases, from
$0.057\pm0.002$ to $0.037\pm0.002$, yet there is still a statistically significant
trend between \h4 and $z$ (12~\textsigma), thanks in part to the increased
sample size (for SAMI, from 135 to 482).
We do not consider the inverse possibility where \mstar at z=0 is systematically
overestimated, because in light of the positive \h4--\mstar
correlation, increasing the cut in \mstar  at $z=0$ would make the relation
stronger and more significant.

To avoid completely the bias inherent to measuring stellar masses across a
long cosmic interval, we can use alternatives: \sigap and dynamical
masses. In Fig.~\ref{f.r.h4evoalt.a} we show quiescent galaxies selected to
have \sigap$> 205$~\kms, roughly corresponding
to the stellar-mass selection $\mstar > 10^{11} \, \mathrm{M}_\odot$
\footnote{The cut in \sigap was derived as the median \sigap for quiescent
galaxies within 0.1~dex from $10^{11}\;\msun$. In the local
Universe, this value corresponds approximately to where early-type galaxies
dominate the velocity-dispersion function \citep{sheth+2003}.
}.
Considering all three surveys, the correlation between \h4 and $z$ has
$\rho = -0.35$ and $P=1.3\times 10^{-20}$ (9-\textsigma significance).
Similarly, in panel~\subref{f.r.h4evoalt.b}, we re-select the sample
based on $M_\mathrm{vir} > 10^{11.5}\,\mathrm{M}_\odot$; in this
case, the correlation coefficient is $\rho=-0.32$, and the probability that
\h4 and $z$ are uncorrelated is $P=8.3\times 10^{-11}$ (6-\textsigma
significance). Similar results can be obtained by swapping
$M_\mathrm{vir}$ with $M_\mathrm{JAM}$, except that the sample size is
smaller (see \S~\ref{s.das.ss.anc}).

\subsection{Measurement bias}\label{s.r.ss.mbias}

Could the reported redshift evolution be due to measurement bias? As we have
seen in \S~\ref{s.das.ss.hok.sss.mbias}, our \h4 measurements depend on the
data homogenisation process. However, the sign of the offset in
Fig.~\ref{f.das.h4bias.c} is such that, had we neglected data homogenisation,
we would have found an even larger \h4 for SAMI galaxies, so an even
\emph{stronger} redshift evolution of \h4.

Overall, we stress that even though the choice of template library affects the
magnitude of the redshift evolution, we find a statistically significant
correlation in every case, with the lowest statistical significance for the
MILES stellar library (only three \textsigma). The different spectral libraries
give systematically different values of \h4, which underscores the challenge
of measuring \h4 in absolute terms, and may complicate comparisons with
numerical simulations.

Another possible source of bias is the use of fixed-size apertures rather than
apertures matching the size of each galaxy. For LEGA-C, given the slit width of
1~arcsec and a galaxy half-light semi-major axis $0.3 < R_\mathrm{e} <
1.7$~arcsec, the LEGA-C half slits span a median fraction of 0.8~$\mathrm{R_e}$,
with the 5\textsuperscript{th}--95\textsuperscript{th} percentiles of the size
distribution spanning between 1.7 to 0.3~\re. The covering fraction
is however larger, because the slits span the full extent of the galaxies along
the spatial direction. Moreover, for SAMI, we compare the measurements inside
the reconstructed LEGA-C slit to the measurements inside one $\mathrm{R_e}$
(Fig.~\ref{f.das.h4bias.a}), finding only a small bias, so we conclude that
aperture effects are not determining the redshift evolution of \h4.

Finally, we note that LEGA-C observations consist of a large number of stacked
exposures \citep[up to 80,][]{vanderwel+2016}, whereas SAMI and MAGPI rely on
up to seven and twelve exposures, respectively. In principle, stacking could
degrade the LOSVD, due to sub-pixel shifts in wavelength and changes to the
atmospheric seeing. However, we find some evidence of redshift evolution even
within LEGA-C alone, which is unlikely to be due to stacking.
From a physical perspective, the reported increase in \h4 with cosmic time is
qualitatively consistent with the previously reported decrease in the
rotation-to-dispersion ratio \citep[$V/\sigma$,][]{newman+2015, toft+2017,
newman+2018, bezanson+2018a}. The latter
is very unlikely to result from stacking - if anything, any degradation of the
LOSVD due to stacking is more likely to decrease $V$ \citep[and increase
$\sigma$,][]{cappellari+2009}, which would cause a spurious \emph{increase}
in $V/\sigma$ with cosmic time, the opposite of what has been reported using 
LEGA-C data \citep{bezanson+2018a}.

\section{Discussion}\label{s.d}

\subsection{Redshift evolution: the role of environment}\label{s.d.ss.enveff}

We find that the average \h4 of massive ($\mstar > 10^{11} \, \msun$) quiescent
galaxies increases with cosmic time (\S~\ref{s.r.ss.h4z}). Looking
at the median \h4 of each survey in Fig.~\ref{f.r.h4evo.a}, it appears that
the evolution happened almost linearly in redshift space between $z=0.8$ and
$z=0.05$.
This is somewhat in tension with simulations, which report that most of the spin down of
galaxies happens after $z=0.5$ \citep{lagos+2018a}. Moreover, by studying the
shape distribution of galaxies, \citet{zhang+2022} also find that the fraction of
massive, disc-like galaxies drops faster between $z=0.15$ and $z=0.45$ than
between $z=0.45$ and $z=0.75$ (cf.~their fig.~6).
Given the small sample size of the MAGPI dataset used in this work, it is unclear
whether our findings are significant or not.
Taking this result at face value, a possible explanation is provided by
environmental effects. Environment is known to correlate with the kinematic structure
of galaxies \citep[e.g.,][]{dressler+1987, cappellari+2011b, deugenio+2015} and MAGPI
samples uniformly in halo mass \citep{foster+2021}, thereby introducing a bias toward
high-density environments. As an example, slow rotators, which are intrinsically round
and dispersion-supported galaxies, are more common in high-density
environments than they are in the field. This is true both in the local
Universe \citep{cappellari+2013b, vandesande+2021b} as well as at the
look-back time of LEGA-C \citep{cole+2020}. It is therefore possible that
part of the evolution (and lack thereof) of \h4 is due to environment
effects which we do not take into account. However, even though we do find a
significant correlation between \h4 and local environment \citetext{for LEGA-C,
we have $P = 5 \times 10^{-4}$, using the local overdensity $\delta$ from
\citealp{sobral+2022}, measured as described in \citealp{darvish+2014},
\citealp{darvish+2016} and \citealp{darvish+2017}}, this correlation is likely
due to the mass-environment correlation.
We can separate the environment and mass dependence of \h4 using partial
correlation coefficients \citep[PCCs; see e.g.,][]{bait+2017, bluck+2019,
baker+2022}. The PCC $\rho(x, y, \vert z)$ measures the correlation coefficient 
of the two random variables $x$ and $y$ while controlling for the third variable
$z$. We find $\rho(\h4, \mstar, \vert \delta) = 0.25$, with $P=2.3\times
10^{-7}$, while $\rho(\h4, \delta , \vert \mstar) = 0.08$, with $P=0.1$. Another
complication is due to the fact that MAGPI galaxies tend to be centrals, which
may be even more likely to have different assembly histories than satellites. A
larger sample may clarify whether environment plays any role.

\subsection{Redshift evolution: beating progenitor bias}\label{s.d.ss.zevo}

We have seen that the population of massive quiescent galaxies has on average
higher \h4 at $z=0.06$ than it had at $z=0.8$. Physically, this means that local
quiescent galaxies have less rotation support and higher radial anisotropy than
quiescent galaxies 7~Gyr ago \citepalias{deugenio+2023}. To disentangle radial anisotropy from
rotation support, we follow the approach presented in \citetalias{deugenio+2023}. If we
consider only round galaxies (observed axis ratio $q \geq 0.8$), we still find
evidence for \h4 evolution (though only 3.5~\textsigma). Alternatively, we
consider only galaxies with low rotation-to-dispersion ratio. Given the
inhomogeneous nature of the \vse measurements for our data \citepalias{deugenio+2023}, we
consider either galaxies with $\vse < 0.5$, or galaxies with \vse less than the
10\textsuperscript{th} percentile of each sample; we obtain statistically
significant results in both cases (5~\textsigma). This underscores that the observed
differences in integrated \h4 reflect not only differences in \vse
\citep[reported in][]{bezanson+2018a}, but also differences in \emph{spatially
resolved} \h4, which is a proxy for radial anisotropy.
Taking our \h4 measurements at face value, and assuming an isothermal
potential \citep{cappellari+2015, poci+2017, derkenne+2021}, the increase from
$\h4\approx0.02$ at $z=0.8$ to $\h4\approx0.06$ at $z=0.06$ means that the radial
orbital anisotropy increases from $\beta\approx0.1$ to $\beta\approx0.4$
\citep[see][their fig.~8]{gerhard1993}. Theoretical expectations,
\citep{vandermarel+franx1993, gerhard1993}, as
well as empirical correlations with \vse and $q$ \citepalias{deugenio+2023}, suggest that
local quiescent galaxies must have rounder intrinsic shapes and lower
rotation-to-dispersion ratios compared to higher-redshift quiescent galaxies.
For this reason, the larger \h4 in low-redshift galaxies is qualitatively
consistent with both the observed differences in shape \citep{vanderwel+2011}
and in rotation support \citep{bezanson+2018a}.

The fact that, at any given redshift, star-forming galaxies have lower \h4,
larger \vse and flatter shapes than coeval quiescent galaxies means that the
changes in the quiescent population cannot be explained by changing
demographics. This is because newly quiescent galaxies, which come necessarily
from the star-forming population, would need to dramatically alter their orbital
structure just before becoming quiescent. Although in principle possible, this
scenario seems unlikely, because --- in the cosmic epochs we are studying ---
most newly quiescent galaxies are expected to become quiescent without
substantial structural change \citep{wu+2018}.
When we attempt to account for progenitor bias (Fig.~\ref{f.r.h4evo.b}), we find
an even stronger redshift evolution, so the
changes in the quiescent population over the last 7~Gyr occur \textit{despite}
the constant influx of newly quiescent galaxies, not because of it.

This is very different from the evolution of e.g. galaxy size, where the
light-weighted size of quiescent galaxies increases because of both
demographics changes and physical processes in individual galaxies
\citep[e.g.][]{vanderwel+2014}. Thus
\h4, and stellar kinematics more in general, enable us to circumvent progenitor
bias. Even though we cannot yet say precisely by how much \h4
increases for the average galaxy, we can firmly establish that individual
quiescent galaxies evolve over cosmic time, even after becoming quiescent.

\subsection{Physical interpretation and future outlook}\label{s.d.ss.fut}

\begin{figure*}
  \includegraphics[type=pdf,ext=.pdf,read=.pdf,width=1.\textwidth]{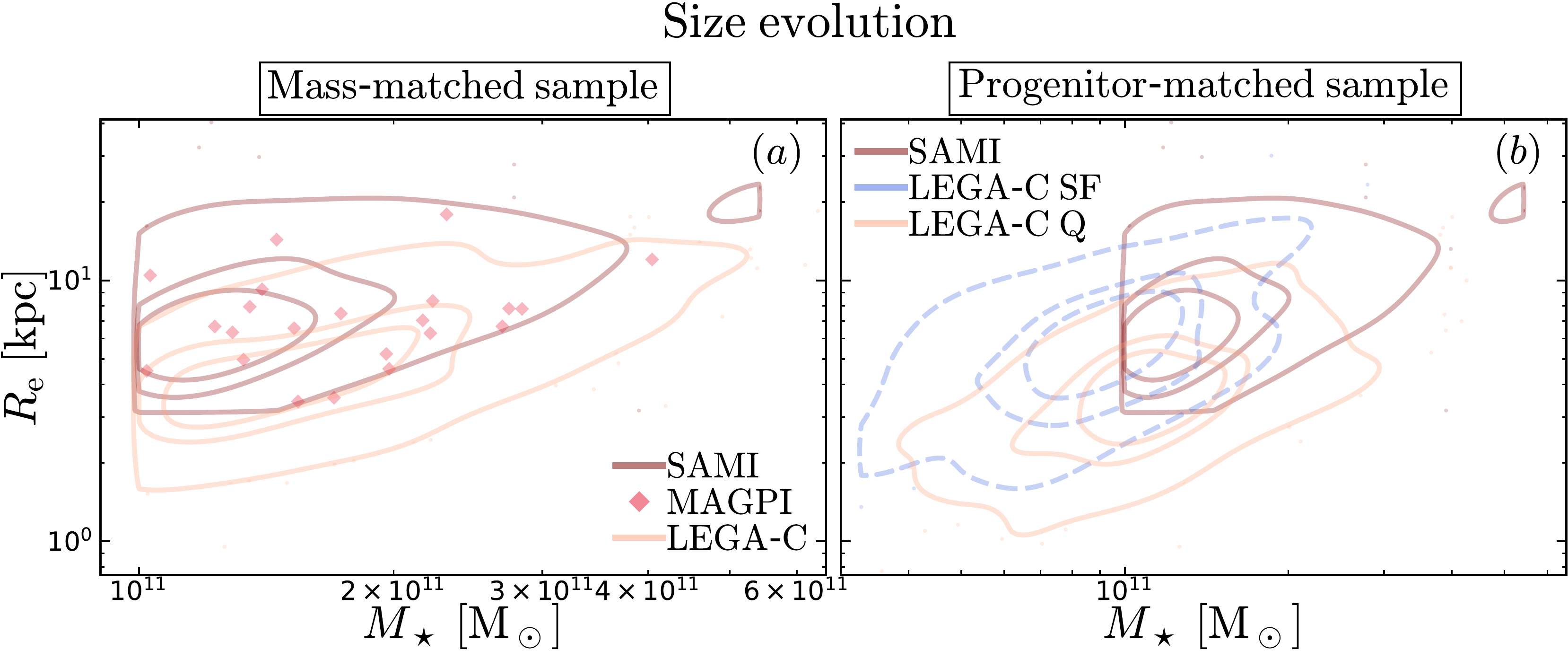}
  {\phantomsubcaption\label{f.d.masssize.a}
   \phantomsubcaption\label{f.d.masssize.b}
  }
  \caption{We find evidence for size evolution for both the mass-matched
  sample (panel~\subref{f.d.masssize.a}) and the progenitor-matched sample
  (panel~\subref{f.d.masssize.b}). The meaning of the symbols is the same
  as Fig.~\ref{f.r.h4evo}.
  }
\end{figure*}

Qualitatively, our results are consistent with the two-phase formation scenario:
around cosmic noon, when gas accretion rates are large, massive galaxies form
through dissipation, giving rise to intrinsically flat, rotation-supported
systems \citep{vanderwel+2011, newman+2018, bezanson+2018a}, characterised by low \h4,
typical of discy, star-forming galaxies. Over time, the accretion rate of cold gas
slows down and is eventually surpassed by the accretion rate of gas-poor
satellites; starved of fuel, star formation starts to decline until the galaxy
becomes quiescent. The takeover of dry mergers means that newly added stars are
distributed on radially-biased orbits, reflecting the infall orbit of their
parent satellite. Collisionless evolution prevents the settling of the accreted
mass on any pre-existing disc, while quiescence prevents the formation of a new
disc. Moreover, the large number of low-mass satellites is critical to explain the
round shape, large size and low rotation support of local massive quiescent galaxies
\citep[][but note that these simulations did not include feedback from
super-massive black holes]{bois+2011, naab+2014}.
The two-phase formation scenario was invoked to explain the observed size growth
of quiescent galaxies. For our mass-matched sample, we find indeed a median
size increase from 4.4 to 7.6~kpc between LEGA-C and SAMI
(Fig.~\ref{f.d.masssize.a}; the uncertainties on these median values are
0.02~dex). This, however, does not take into account progenitor bias. For the
progenitor-matched sample (Fig.~\ref{f.d.masssize.b}), the median size of the
quiescent progenitors is 3.8~kpc, the median size of the star-forming progenitors
is 5.7~kpc. The probability that the samples match in their \re distributions are
all $P_\mathrm{KS}<10^{-5}$.
We cannot exclude secular processes as being partly responsible for the observed
evolution in \h4, but it is very unlikely that they are the only process. First,
because these processes can only re-distribute angular momentum; the transition
from high-angular-momentum, low-\h4 galaxies to low-angular-momentum , high-\h4
galaxies requires external processes. Moreover, internal processes alone cannot
explain the size growth of quiescent galaxies.

\label{f.d.masssize}

We already know that --- for stellar kinematics --- the degree of rotation support
decreases with cosmic time \citep{bezanson+2018a}. Our results are consistent
with their findings, but add that, in parallel with the decrease in rotation
support, there is an increase in radial anisotropy, which is expected from
minor dry mergers. Note that single major dry merger events may
not necessarily increase the fraction of radial orbits \citep{bois+2011}.
In contrast, numerous dry mergers can even out the asymmetry of individual
satellite orbits, by averaging over a sufficient number of orbits.

Incidentally, our
results also reconcile the observation that, for stars, the rotation-to-dispersion
ratio decreases with cosmic time \citep{bezanson+2018a}, but for star-forming gas
it increases with cosmic time \citep[e.g.,][]{leroy+2009, law+2009, schreiber+2009,
wisnioski+2015}. These opposite trends raise the question of how is it possible
that the dispersion of stars increases with time, but the dispersion of newly
formed stars decreases with cosmic time. The answer is that, even though in-situ
stars form on progressively thinner discs\footnote{This may be due to the fact
that, as time passes, the gravitational potential of massive galaxies becomes more
centrally concentrated, stabilising the gas disc \citep{hopkins+2023}.}, the
evolution of the quiescent population is driven by
accretion of ex-situ stars on radially anisotropic orbits \citep{lagos+2017}. In
the mass regime we are probing, these dry mergers overcome the demographics
change due to the addition of newly quiescent galaxies that had thinner discs
than the previous generations of long-quiescent galaxies.

A clear prediction of the proposed scenario is that \h4 must be linked to the
fraction of \textit{ex-situ} stars, which can be readily identified in numerical
simulations, or, alternatively, could be estimated using a joint chemo-dynamical
analysis \citep[see e.g.][]{poci+2019}.
Given that our measurements do not require spatially resolved spectroscopy, there is
a lot we can learn from large single-fibre surveys such as the Sloan Digital Sky
Survey \citep{york+2000}. The current generation of large single-fibre surveys of the
local Universe will give us access to even larger samples \citetext{e.g. the
4MOST Hemisphere Survey, Taylor et~al., in~prep.; the WEAVE-StePS,
\citealp{costantin+2019} and 4MOST-StePS surveys; and the DESI Bright
Galaxy Survey, \citealp{ruiz-macias+2021}}, while future high-redshift surveys
will enable us to study \h4 for galaxies at cosmic noon
\citep[MOONRISE survey,][]{maiolino+2020}.

\subsubsection{Caveats and limitations}\label{s.d.ss.calim}

There are however some difficulties with our interpretation. First, the inferred
evolution comes mostly from a single dataset: the LEGA-C survey; using only SAMI
and MAGPI, we would not find evidence of evolution.
Stacking data from higher redshift surveys \citep[e.g., the
VIRIAL survey,][]{mendel+2015}, or larger datasets from future surveys
\citep[e.g., MOONRISE,][]{maiolino+2020}, could address this shortcoming.

Second, the bulk of the evolution due to low-mass dry mergers is expected to
occur before $z=1$
\citep[e.g.][]{remus+2013, remus+2017, springel+2018, karademir+2019}. Evolution
between $z=0.8$ and $z=0.06$ seems unlikely in this context \citep[but see][for a
different view]{oser+2012, peirani+2019}. The fact
that we find a
strong correlation between \h4 and stellar surface mass density, even when no
correlation with \mstar is found (Appendix~\ref{app.h4snr}), suggests a link with
the shape of the mass profile. Mass redistribution after feedback due to
super-massive black holes, or after major dry mergers, could be responsible for
reorganising the orbital structure of massive galaxies.

Finally, we note that, after a preliminary analysis from the Magneticum
simulations \citep{hirschmann+2014, teklu+2015}, we find evidence for a slight
increase of \h4 over the redshift range considered here (Remus et al., in~prep.).
A possible explanation could be that \h4 inferred from
simulations and observations are not immediately comparable, partly because of
our unique observing setup, partly because of the difficulty of measuring \h4 in
an absolute sense \citepalias{deugenio+2023}.

\section{Conclusions}\label{s.c}

In this work, we studied the cosmic evolution of stellar kinematics in high-mass
quiescent galaxies. We used the parameter \h4, the coefficient of the
4\textsuperscript{th}-order Hermite polynomial in the Gauss-Hermite expansion of
the line-of-sight velocity distribution. These empirical measurements are dominated
by systematic errors, which we minimise by homogenising the data.
The combination of the SAMI, MAGPI and
LEGA-C surveys enables us to leverage a long baseline in cosmic time, covering
more than half the history of the Universe.
\begin{enumerate}
  \item using a mass-matched sample, we find strong evidence of redshift
  evolution for integrated \h4 (7~\textsigma). The median value of \h4 increases
  from $0.019\pm0.002$ at $z=0.82$, to $0.045\pm0.008$ at $z=0.32$ up to
  $0.059\pm0.004$ at $z=0.06$.
  \item The reported redshift evolution is robust against progenitor bias:
  using a `progenitor-matched' sample, which includes lower-mass quiescent and
  star-forming galaxies in the high-redshift bin, the inferred evolution of \h4
  becomes even stronger.
  \item The reported evolution suggests an increase in the light-weighted radial
  anisotropy, which is consistent with the outcome of accretion of gas-poor satellites;
  however, other interpretations are also possible (\S~\ref{s.d.ss.calim}).
\end{enumerate}
Future observations will be needed to independently check the reported evolution, while
forward modelling from simulated galaxies may help compare the reported evolution with
theoretical predictions.

\section*{Acknowledgements}

We thank the anonymous referee for insightful comments that greatly improved this article.
FDE and AvdW acknowledge funding through the H2020 ERC Consolidator Grant 683184.
FDE and RM acknowledge funding through the ERC Advanced grant 695671 ``QUENCH'' and
support by the Science and Technology Facilities Council (STFC).
SB acknowledges funding support from the Australian Research Council through a Future Fellowship (FT140101166) ORCID - 0000-0002-9796-1363.
CF is the recipient of an Australian Research Council Future Fellowship
(project number FT21010016) funded by the Australian Government.
H{\"U} gratefully acknowledges support by the Isaac Newton Trust and by the Kavli Foundation.
LMV acknowledges support by the COMPLEX project from the European Research Council (ERC) under the European Union’s Horizon 2020 research and innovation program grant agreement ERC-2019-AdG 882679.
through a Newton-Kavli Junior Fellowship.
JvdS acknowledges support of an Australian Research Council Discovery Early Career Research Award (project number DE200100461) funded by the Australian Government. 
JJB acknowledges support of an Australian Research Council Future Fellowship (FT180100231).

The SAMI Galaxy Survey is based on observations made at the Anglo-Australian
Telescope. The Sydney-AAO Multi-object Integral-field spectrograph (SAMI) was
developed jointly by the University of Sydney and the Australian Astronomical
Observatory, and funded by ARC grants FF0776384 (Bland-Hawthorn) and
LE130100198. The SAMI input catalog is based on data taken from the Sloan
Digital Sky Survey, the GAMA Survey and the VST ATLAS Survey. The SAMI Galaxy
Survey is funded by the Australian Research Council Centre of Excellence for
All-sky Astrophysics (CAASTRO), through project number CE110001020, and other
participating institutions. The SAMI Galaxy Survey website is
http://sami-survey.org/ .

Funding for SDSS-III has been provided by the Alfred P. Sloan Foundation, the
Participating Institutions, the National Science Foundation, and the U.S.
Department of Energy Office of Science. The SDSS-III web site is
http://www.sdss3.org/ .

GAMA is a joint European-Australasian project based around a spectroscopic
campaign using the Anglo-Australian Telescope. The GAMA input catalogue is
based on data taken from the Sloan Digital Sky Survey and the UKIRT Infrared
Deep Sky Survey. Complementary imaging of the GAMA regions is being obtained
by a number of independent survey programmes including GALEX MIS, VST KiDS,
VISTA VIKING, WISE, Herschel-ATLAS, GMRT and ASKAP providing UV to radio
coverage. GAMA is funded by the STFC (UK), the ARC (Australia), the AAO, and
the participating institutions. The GAMA website is http://www.gama-survey.org/ .

Based on observations made with ESO Telescopes at the La Silla Paranal Observatory
under programme IDs 179.A-2004. Based on observations collected at the European
Organisation for Astronomical Research in the Southern Hemisphere under ESO
programme 1104.B-0536. We wish to thank the ESO staff, and in particular the
staff at Paranal Observatory, for carrying out the MAGPI observations. Part of this 
research was conducted by the Australian Research Council Centre of Excellence
for All Sky Astrophysics in 3 Dimensions (ASTRO 3D), through project number
CE170100013.

This work made extensive use of the freely available
\href{http://www.debian.org}{Debian GNU/Linux} operative system. We used the
\href{http://www.python.org}{Python} programming language
\citep{vanrossum1995}, maintained and distributed by the Python Software
Foundation. We further acknowledge direct use of
{\sc \href{https://pypi.org/project/astropy/}{astropy}} \citep{astropyco+2013},
{\sc \href{https://pypi.org/project/matplotlib/}{matplotlib}} \citep{hunter2007},
{\sc \href{https://pypi.org/project/numpy/}{numpy}} \citep{harris+2020},
{\sc \href{https://pypi.org/project/pathos/}{pathos}} \citep{mckerns+2011},
{\sc \href{https://pypi.org/project/pingouin/}{pingouin}} \citep{vallat2018},
\href{https://pypi.org/project/ppxf/}{\ppxf} \citep{cappellari2017},
{\sc \href{https://github.com/asgr/ProSpect}{prospect}} \citep{robotham+2020},
{\sc \href{https://pypi.org/project/scipy/}{scipy}} \citep{jones+2001}
and {\sc \href{http://www.star.bris.ac.uk/~mbt/topcat/}{topcat}} \citep{taylor2005}.

\section*{Data availability}

The reduced data used in this work is available in the public domain. For SAMI,
through the \href{https://docs.datacentral.org.au/sami}{SAMI Data Release 3}
\citep{croom+2021a}. Ancillary
data come from the \href{http://gama-survey.org}{GAMA Data Release 3}
\citep{baldry+2018} and raw data are from
\href{https://classic.sdss.org/dr7/}{SDSS DR7} \citep{abazajian+2009},
\href{https://www.sdss3.org/dr9/}{SDSS DR9} \citep{ahn+2012} and
\href{http://casu.ast.cam.ac.uk/vstsp/imgquery/search}{VST}
\citep{shanks+2013, shanks+2015}.
For MAGPI, the raw data (and a basic data reduction) are available through the
the \href{http://archive.eso.org/cms.html}{ESO Science Archive Facility}.
For LEGA-C, the raw data and a catalogue of basic photometric and kinematic
measurements are available through the
\href{http://archive.eso.org/cms.html}{ESO Science Archive Facility}.

Integrated \h4 measurements are available
\sendemail{francesco.deugenio@gmail.com}{h4 evolution: data request}{
contacting the corresponding author.}

\section*{Affiliations}
\noindent
{\it
\hypertarget{aff11}{$^{11}$}Australian Astronomical Optics, Astralis-USydney, School of Physics, University of Sydney, NSW 2006, Australia\\
\hypertarget{aff12}{$^{12}$}Research School of Astronomy and Astrophysics, Australian National University, Canberra, ACT 2611, Australia\\
\hypertarget{aff13}{$^{13}$}International Centre for Radio Astronomy Research (ICRAR), University of Western Australia, Crawley, WA 6009, Australia\\
\hypertarget{aff14}{$^{14}$}Research Centre for Astronomy, Astrophysics and Astrophotonics, School of Mathematical and Physical Sciences, Macquarie University, Sydney, NSW 2109, Australia\\
\hypertarget{aff15}{$^{15}$}Astronomy Department, Yale University, New Haven, CT 06511, USA\\
\hypertarget{aff12}{$^{16}$}INAF-Osservatorio Astrofisico di Arcetri, Largo Enrico Fermi 5, I-50125 Firenze, Italy\\
\hypertarget{aff17}{$^{17}$}Leiden Observatory, Leiden University, P.O. Box 9513, 2300 RA, Leiden, The Netherlands\\
\hypertarget{aff18}{$^{18}$}Max-Planck-Institut f\"ur Astronomie, K\"onigstuhl 17, D-69117, Heidelberg, Germany\\
\hypertarget{aff19}{$^{19}$}Department of Physics and Astronomy, University College London, Gower Street, London WC1E 6BT, UK\\
\hypertarget{aff20}{$^{20}$}Department of Astronomy, University of Wisconsin, 475 N. Charter Street, Madison, WI 53706, USA\\
\hypertarget{aff21}{$^{21}$}Space Telescope Science Institute, 3700 San Martin Drive, Baltimore, MD 21218, USA\\
\hypertarget{aff22}{$^{22}$}Centre for Extragalactic Astronomy, University of Durham, Stockton Road, Durham DH1 3LE, United Kingdom\\
\hypertarget{aff23}{$^{23}$}Universit\"ats-Sternwarte, Fakult\"at f\"ur Physik, Ludwig-Maximilians-Universit\"at M\"unchen, Scheinerstr. 1, 81679 M\"unchen, Germany\\
\hypertarget{aff24}{$^{24}$}School of Mathematics and Physics, University of Queensland, Brisbane, QLD 4072, Australia\\
\hypertarget{aff25}{$^{25}$}Department of Astrophysics, University of Vienna, T\"urkenschanzstra{\ss}e 17, 1180 Vienna, Austria\\
\hypertarget{aff26}{$^{26}$}George P. and Cynthia Woods Mitchell Institute for Fundamental Physics and Astronomy, Texas A\&M University, College Station, TX 77843-4242, USA\\
}

%%%%%%%%%%%%%%%%%%%% REFERENCES %%%%%%%%%%%%%%%%%%

% The best way to enter references is to use BibTeX:

\bibliographystyle{mnras}
\bibliography{h4evo.bbl}

%%%%%%%%%%%%%%%%%%%%%%%%%%%%%%%%%%%%%%%%%%%%%%%%%%

%%%%%%%%%%%%%%%%% APPENDICES %%%%%%%%%%%%%%%%%%%%%

\appendix

\section{Mass and \texorpdfstring{$S/N$}{S/N} correlations}\label{app.h4snr}

\begin{figure*}
  \includegraphics[type=pdf,ext=.pdf,read=.pdf,width=1.\textwidth]{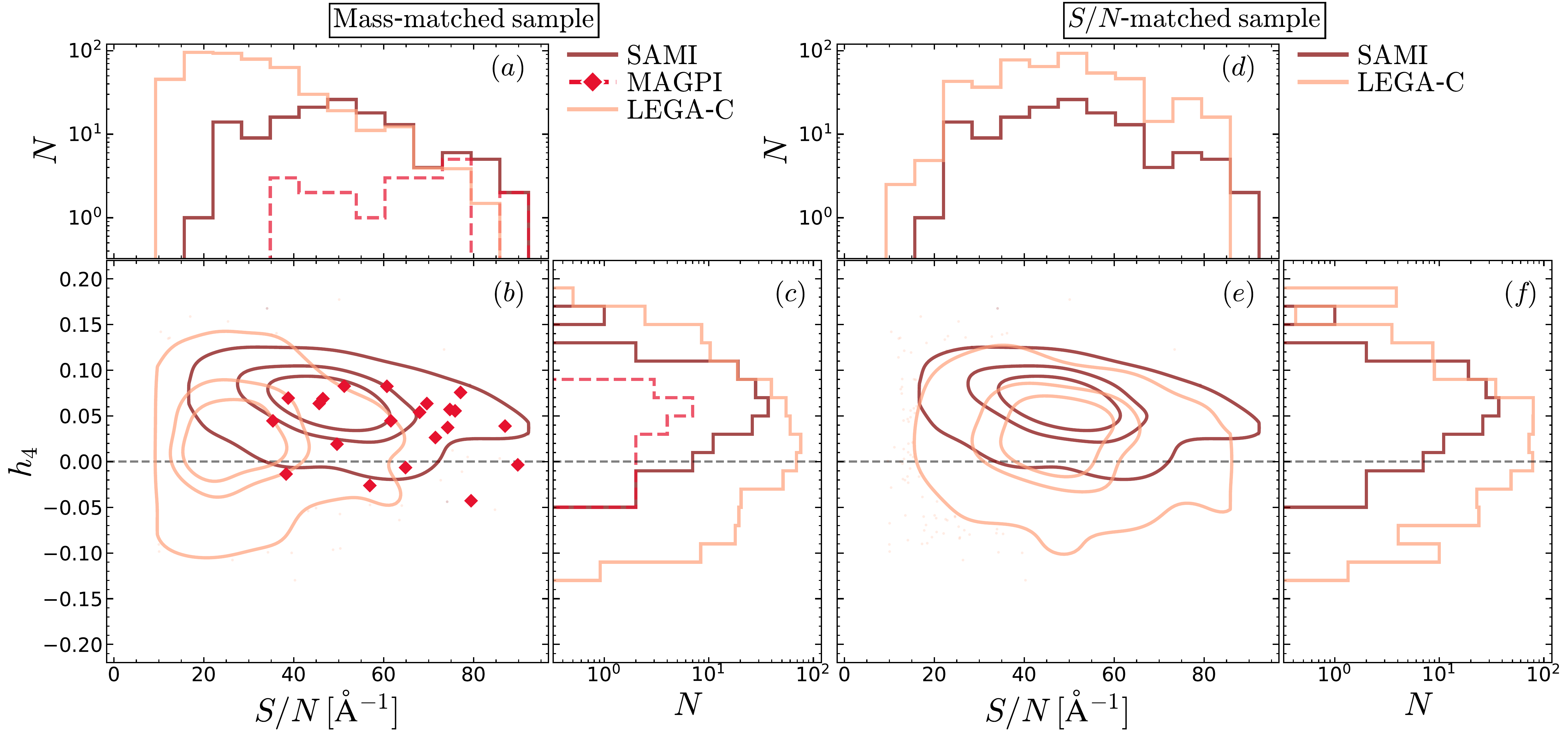}
  {\phantomsubcaption\label{f.app.h4snr.a}
   \phantomsubcaption\label{f.app.h4snr.b}
   \phantomsubcaption\label{f.app.h4snr.c}
   \phantomsubcaption\label{f.app.h4snr.d}
   \phantomsubcaption\label{f.app.h4snr.e}
   \phantomsubcaption\label{f.app.h4snr.f}}
  \caption{$S/N$ and \h4 distribution of the mass-matched sample 
  (panels~\subref{f.app.h4snr.a}--\subref{f.app.h4snr.c}) and for a $S/N$-matched
  sample (panels~\subref{f.app.h4snr.d}--\subref{f.app.h4snr.f}). The differences
  in $S/N$ between SAMI and LEGA-C (panel~\subref{f.app.h4snr.a}) do not
  explain the observed difference in \h4 (panel~\subref{f.app.h4snr.c}), because
  the latter persists even after matching the two samples in $S/N$
  (panel~\subref{f.app.h4snr.d}).
  }\label{f.app.h4snr}
\end{figure*}

In Fig.~\ref{f.app.h4snr} we show the relation between \h4 and $S/N$,
for the mass-matched sample
(panels~\subref{f.app.h4snr.a}--\subref{f.app.h4snr.c}) and for a $S/N$-matched
sample of quiescent galaxies (panels~\subref{f.app.h4snr.d}--\subref{f.app.h4snr.f}).

In panel~\subref{f.app.h4snr.a} we show that the three samples have different
$S/N$ distributions. This seems concerning due to the way we measure \h4: \ppxf
features a built-in penalisation against non-Gaussian solutions, so that \h4
from low-$S/N$ spectra is biased towards \h4=0. However, this is not what causes
the different \h4 between the different samples, because repeating our analysis
without penalisation gives statistically consistent results.

In \citetalias{deugenio+2023}, we reported no independent correlation between $S/N$ and \h4;
here, we find a significant anti-correlation for SAMI ($\rho = -0.30$,
$P=0.0004$), marginal evidence of an anti-correlation for MAGPI ($\rho = -0.29$,
$P=0.19$), and no correlation for LEGA-C ($\rho = 0.002$, $P=0.96$). As for the
\h4--\mstar correlation (\S~\ref{s.r.ss.h4z}), the difference between our results
and \citetalias{deugenio+2023} is due to different sample selection. Using the same mass cut as
\citetalias{deugenio+2023}, the \h4--$S/N$ correlation has $\rho = 0.10$ and $P = 0.03$
It is unclear why, in the \mstar range considered here, we find a strong
h4--$S/N$ anti-correlation. We propose that it arises from an even stronger
stronger anti-correlation between \h4 and central surface mass density. We test
this hypothesis using again partial correlation coefficients (see
\S~\ref{s.d.ss.enveff}). For SAMI, we find
$\rho(h_4, \Sigma_\star(R<R_\mathrm{e}) \vert S/N) = -0.31$, with $P=3\times
10^{-4}$; in contrast, $\rho(h_4, S/N \vert \Sigma_\star(R<R_\mathrm{e})) =
-0.26$, with $P=2 \times 10^{-3}$.

In panels~\subref{f.app.h4snr.d}--\subref{f.app.h4snr.f} we compare the \h4 and
$S/N$ distributions of the SAMI quiescent sample, to a subset of the LEGA-C
quiescent sample, chosen to match the $S/N$ distribution of SAMI
(panel~\subref{f.app.h4snr.d}). Despite having the same $S/N$ by construction,
the two samples have different \h4 distributions (panel~\subref{f.app.h4snr.f}),
underscoring that the difference in \h4 between SAMI and LEGA-C cannot be
explained by different $S/N$ levels.

%%%%%%%%%%%%%%%%%%%%%%%%%%%%%%%%%%%%%%%%%%%%%%%%%%

% Don't change these lines
\bsp	% typesetting comment
\label{lastpage}
\end{document}